\documentclass[times,twocolumn,final]{elsarticle}
\usepackage{medima}
\usepackage{framed,multirow}

\usepackage{amssymb}
\usepackage{latexsym}

\usepackage{url}
\usepackage{xcolor}

\usepackage{hyperref}

\usepackage{multirow}
\usepackage{bbm}
\usepackage{bm}
\usepackage{physics}
\usepackage[normalem]{ulem}
\useunder{\uline}{\ul}{}
\usepackage{adjustbox}

\definecolor{newcolor}{rgb}{.8,.349,.1}

\journal{Medical Image Analysis}

\begin{document}

\verso{Yi Gu \textit{et~al.}}

\begin{frontmatter}

\title{Bone mineral density estimation from a plain X-ray image by learning decomposition into projections of bone-segmented computed tomography}

\author[1,4]{Yi \snm{Gu}\corref{cor1}}
\ead{gu.yi.gu4@is.naist.jp}
\cortext[cor1]{Corresponding authors at ICB lab., Division of Information Science, Graduate School of Science and Technology, Nara Institute of Science and Technology, Japan.}
\author[1]{Yoshito \snm{Otake}\corref{cor1}}
\ead{otake@is.naist.jp}

\author[2]{Keisuke \snm{Uemura}}

\author[1]{Mazen \snm{Soufi}}

\author[3]{Masaki \snm{Takao}}

\author[4]{Hugues \snm{Talbot}}

\author[2]{Seiji \snm{Okada}}

\author[5]{Nobuhiko \snm{Sugano}}

\author[1]{Yoshinobu \snm{Sato}\corref{cor1}}
\ead{yoshi@is.naist.jp}

% \address[1]{Affiliation 1, Address, City and Postal Code, Country}
\address[1]{Graduate School of Science and Technology, Nara Institute of Science and Technology, Ikoma, Nara 630-0192, Japan}
\address[2]{Department of Orthopaedic Surgery, Osaka University Graduate School of Medicine, Suita, Osaka 565-0871, Japan}
\address[3]{Department of Bone and Joint Surgery, Ehime University Graduate School of Medicine, Toon, Ehime 791-0295, Japan}
\address[4]{CentraleSupélec, Université Paris-Saclay, Inria, Gif-sur-Yvette 91190, France}
\address[5]{Department of Orthopedic Medical Engineering, Osaka University Graduate School of Medicine, Suita, Osaka 565-0871, Japan}

% \received{1 May 2013}
% \finalform{10 May 2013}
% \accepted{13 May 2013}
% \availableonline{15 May 2013}
% \communicated{S. Sarkar}

\begin{abstract}
%\newcommand{\rmtxt}[1]{\textcolor{red}{\sout{#1}}}
%\newcommand{\adtxt}[1]{\textcolor{blue}{#1}}
%%%
Osteoporosis is a prevalent bone disease that causes fractures in fragile bones, leading to a decline in daily living activities. 
Dual-energy X-ray absorptiometry (DXA) and quantitative computed tomography (QCT) are highly accurate for diagnosing osteoporosis; however, these modalities require special equipment and scan protocols. 
To frequently monitor bone health, low-cost, low-dose, and ubiquitously available diagnostic methods are highly anticipated. 
In this study, we aim to perform bone mineral density (BMD) estimation from a plain X-ray image for opportunistic screening, which is potentially useful for early diagnosis. 
Existing methods have used multi-stage approaches consisting of extraction of the region of interest and simple regression to estimate BMD, which require a large amount of training data. 
Therefore, we propose an efficient method that learns decomposition into projections of bone-segmented QCT for BMD estimation under limited datasets. 
The proposed method achieved high accuracy in BMD estimation, where Pearson correlation coefficients of 0.880 and 0.920 were observed for DXA-measured BMD and QCT-measured BMD estimation tasks, respectively, and the root mean square of the coefficient of variation values were 3.27 to 3.79\% for four measurements with different poses. 
Furthermore, we conducted extensive validation experiments, including multi-pose, uncalibrated-CT, and compression experiments toward actual application in routine clinical practice.

%%%%
\end{abstract}


\end{frontmatter}

%\linenumbers

\section{Introduction}
%\newcommand{\rmtxt}[1]{\textcolor{red}{\sout{#1}}}
%\newcommand{\adtxt}[1]{\textcolor{blue}{#1}}
% Clinical Background
In the last decades, a significant increase in the prevalence of osteoporosis has been observed. It has become the most common metabolic bone disease and is characterized by a continuous loss of bone mass \citep{compston_osteoporosis_2019,yang_road_2020,zou_advances_2020,noh_molecular_2020,yu_epidemiology_2019}. It affects mostly the elderly population but sometimes affects children also\citep{ward_contemporary_2020}.
The measurement of bone mineral density (BMD), a parameter describing bone strength, is essential for the diagnosis, treatment, and drug development of osteoporosis \citep{loffler_X-ray-based_2020,ward_contemporary_2020,li_osteoporosis_2021,lorentzon_treating_2019,kung_factors_2013,iki_bone_2001}.
In routine clinical practice, dual-energy X-ray absorptiometry (DXA) \citep{iki_bone_2001,mazess_dual-energy_1990,blake_role_2007,kroger_bone_1992,omalley_trends_2011,pisani_screening_2013} is considered the gold standard for BMD measurement.
In addition, quantitative computed tomography (QCT) \citep{engelke_clinical_2008,mueller_phantom-less_2011,aggarwal_opportunistic_2021,ziemlewicz_opportunistic_2015} has been explored with a view to provide opportunistic screening.
Despite the prevalence of osteoporosis worldwide, the diagnosis rates for adults remain unacceptably low in many regions \citep{snodgrass_osteoporosis_2022,choi_prevalence_2012,papaioannou_osteoporosis_2008,mccloskey_osteoporosis_2021}.
Therefore, there is a strong demand for developing more straightforward methods of measuring BMD, providing opportunistic screening for the early detection of osteoporosis, an asymptomatic disease unless a fracture happens.

% \\Solution Trends
Recent studies have focused on the use of deep learning to estimate BMD or diagnose osteoporosis from plain X-ray images \citep{hsieh_automated_2021,yamamoto_deep_2020,jang_prediction_2021,ho_application_2021,gu_bmd-gan_2022}\citep{wang_lumbar_2023}, which are more widespread modalities than DXA and QCT. 
% Most existing methods performed regression to estimate BMD or classification to diagnose osteoporosis \citep{hsieh_automated_2021,yamamoto_deep_2020,jang_prediction_2021,ho_application_2021}, grading from a region of interest (ROI) of X-ray images obtained by manual or automatic localization, some of which achieved high correlations with DXA-measured BMD, and grading using a large-scale training dataset.
Most existing methods directly regress BMD or classify osteoporosis from regions of interest (ROI) of X-ray images, some of which achieved high correlations with DXA-measured BMD \citep{hsieh_automated_2021,wang_lumbar_2023}.
However, these methods required large-scale training datasets and did not provide the spatial density distribution of the target bone.
Furthermore, they did not leverage information from CT.
The critical requirement of a large-scale training database would limit the application of these methods when the target bone of BMD measurement extends to other anatomical areas (e.g., different positions of the vertebrae, pelvis, sacrum, etc.).

% \\Technical Background
Different from end-to-end regression, \citet{whitmarsh_reconstructing_2011} proposed a method of estimating 3D distributions of QCT-measured BMD from DXA by reconstructing 3D shapes and BMD distributions with statistical shape modeling, improving the diagnosis of osteoporosis and fracture risk assessment. 
From the viewpoint of X-ray image processing, bone suppression is one of the main topics \citep{suzuki_image-processing_2006,yang_cascade_2017,liu_automatic_2019,eslami_image--images_2020}, enhancing the visibility of other soft tissues to increase the diagnosis rate of soft-tissue diseases by machines and clinicians. 
These studies used convolutional neural networks to suppress bones \citep{suzuki_image-processing_2006,yang_cascade_2017} or decompose into target soft tissues directly \citep{eslami_image--images_2020}, some of which utilized generative adversarial networks (GANs) \citep{isola_image--image_2017} for their realistic synthesis abilities. 
Because those studies focused on the diagnosis of soft tissues, they did not address the quantitative evaluation of bone decomposition for BMD estimation.
While those studies inspired this one, we propose to learn X-ray image decomposition into projections of bone-segmented QCT for BMD estimation, targeting both DXA-measured BMD (hereafter ``DXA-BMD'') and QCT-measured BMD (hereafter ``QCT-BMD''). 

\begin{figure}[!t]
\centering
\includegraphics[width=\columnwidth]{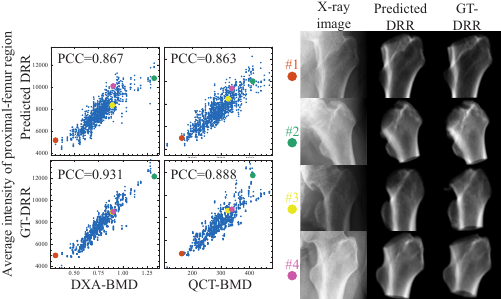}
\caption{Relationship between the intensities of the proximal-femur region of the DRR and BMD values in datasets of 305 patients.
(Left) scatter plots showing the correlation of the average intensity of ground-truth (GT) DRR and predicted DRR with DXA-BMD and QCT-BMD evaluated by Pearson correlation coefficient.
(Right) proximal femur ROIs of five representative cases.
ROIs \#1 and \#2 have similar X-ray intensities but significantly different BMDs, whereas ROIs \#3 and \#4 have similar BMDs but significantly different X-ray intensities. The predicted DRRs correctly recovered the intensity of QCT DRR, regardless of the intensity of the input X-ray image.}
\label{fig:average_intensity}
\end{figure}

\begin{figure*}[!t]
\centering
\includegraphics[scale=1]{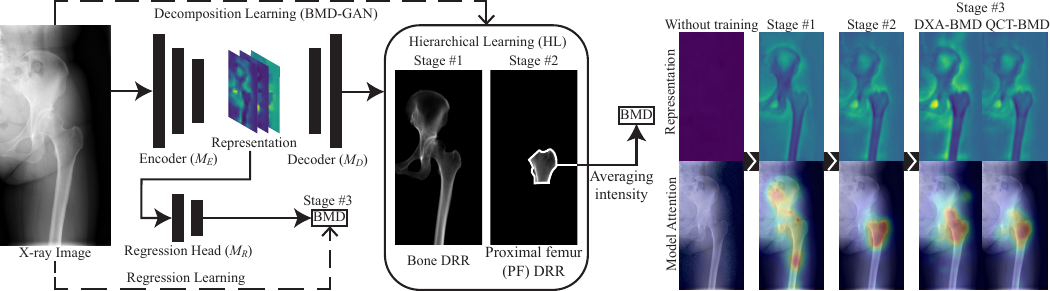}
\caption{(Left) Overview of the proposed method consisting of three training stages, where the model first learns X-ray image representation by decomposition training in stage- one and stage- two with HL applied \citep{gu_bmd-gan_2022} and then learns mapping to BMD by regressing from the learned representation. 
(Right) The visualized procedure of shaping representation and model attention shifting with learning stages.
The representations were visualized after being compressed to one-channel images using principal component analysis; the model attention maps were visualized using FullGrad \citep{srinivas_full-gradient_2019}}
\label{fig:method_overview}
\end{figure*}
% Replace names with symbols

% Concept of the proposed method
Unlike existing BMD estimation methods, we virtually and automatically mimicked the decomposition procedure (similar to the DXA scanning) and the measurement procedure, which is physically and manually conducted in DXA-based BMD measurement, by fully utilizing rich information from QCT in its training phase by predicting bone-segmented digitally reconstructed radiography (DRR), which are projections from segmented QCT. 
Using GAN, we trained a model to decompose an X-ray image into the density distribution of a partial bone defined clinically for BMD measurement. 
Despite the difficulty in GAN training, recent studies \citep{zhao_differentiable_2020,zhang_consistency_2020,wu_gradient_2021,tseng_regularizing_2021} were able to stabilize the training and reduce the demand for a large-scale database. 
For similar purposes, we introduced the hierarchical learning (HL) method so that the small-area regions of the target bone required for BMD estimation were decomposed accurately and stably even without a large training dataset.
Fig. \ref{fig:average_intensity} illustrates the relationship between X-ray images and DXA-BMDs in our patient dataset, demonstrating the challenge in the task of BMD prediction based on X-ray images, which showed no correlation with BMD. 
The proposed method, which disentangled soft tissues and bone from an X-ray image, accurately predicted the target bone, showing a high correlation with BMD.
Furthermore, we showed improvement by regressing BMD from the learned representation generated by the encoder of the decomposition model, demonstrating the effectiveness of the proposed decomposition learning.

% Contributions
We summarize our contributions as follows:
\begin{itemize}
    \item Proposal of a BMD estimation method through X-ray image decomposition.
    \item Proposal of a hierarchical learning framework for decomposition into a small target.
    \item Proposal of a representation learning method for BMD estimation using decomposition learning.
    \item Extensive validation with real clinical datasets involving variations in patient poses, image compression ratios, and CT calibration.
\end{itemize}

% Extension Clarification
This paper is built upon our conference paper \citep{gu_bmd-gan_2022} with extensions into several aspects. 
From the methodological aspect, we explicitly targeted QCT-BMD in addition to DXA-BMD. We conducted an ablation study on thresholds for obtaining binarized masks when calculating BMDs from the predicted DRR. We validated the effectiveness of the proposed decomposition learning by regressing BMD from the learned representation with a transformer-based regression network.
For the dataset aspect, we included more patients who had undergone X-ray scanning and QCT scanning and had taken BMD measurements, resulting in 305 patients (increased from 200 patients) in \citet{gu_bmd-gan_2022}. We included more X-ray images imaged with supine, abduction, and adduction poses in addition to the standing pose.
For the experimental aspect, we evaluated the performance of the proposed method using the extended dataset and conducted additional validation for robustness against pose variation, image compression, and uncalibrated CT. We also investigated error sources by identifying outliers.

\begin{figure*}[!t]
\centering
\includegraphics[scale=1]{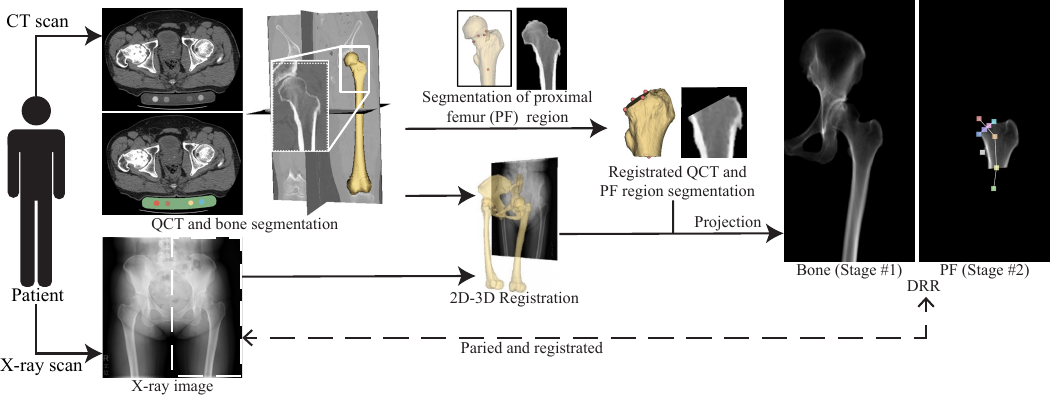}
\caption{Construction of the training dataset consisting of the intensity calibration of CT \citep{uemura_automated_2021}, bone segmentation \citep{hiasa_automated_2020}, 2D-3D registration to the X-ray image \citep{otake_intraoperative_2012} and DRR generation by projecting QCT.}
\label{fig:dataset_const}
\end{figure*}

\section{Related Work}
\subsection{Grading osteoporosis from X-ray images}
\citet{jang_prediction_2021} used a VGG16 model with integrated Non-local Neural Networks to classify osteoporosis from a manually cropped X-ray image.
\citet{hsieh_automated_2021} and \citet{ho_application_2021} developed BMD estimation frameworks using localization and regression models, requiring large-scale databases to achieve high accuracy.
\citet{wang_lumbar_2023} proposed a Transformer-based model for BMD estimation from chest X-ray images with multiple ROIs, simultaneously capturing local and global information to improve accuracy.
Although conventional methods strove to localize the target regions from an X-ray image (which lies in 2D space), they did not address the entanglement of bone and soft tissues (which align in 3D space).
We argue the unsolved tissue entanglement, proposing the utilization of CT information during training in order to allow the estimation of the spatial density of bone from a plain X-ray image.
\citet{yamamoto_deep_2020} proposed an ensemble model combining multiple convolutional neural networks, trying to classify osteoporosis with patients' clinical covariates that are fused in the last layer of the model.
Though most clinical covariates are easy to obtain in routine clinical practice and reasonably improve prediction accuracy, we do not use patient characteristics in our method for better clarification of the improvement.

\subsection{Representation learning}
Representation learning aims to learn efficient representations of data whose information is easier to be extracted for downstream tasks, improving effective learning \citep{bengio_representation_2013}.
Many representation learning methods have been proposed for solving tasks in the fields of natural language processing and general image processing.
\citep{devlin_bert_2019,noroozi_representation_2017,chen_exploring_2021,donahue_large_2019}. However, those methods usually rely on large-scale pretraining (which is hard to achieve using medical images) and they did not utilize the specialty such as penetrability in medical images.
\citet{zhou_latent_2021} proposed a representation method for brain tumor segmentation with missing modes of magnetic resonance images, utilizing the similarity between multi-modal MR images.
\citet{liu_incomplete_2021} proposed a multi-view missing data completion framework to learn multi-modal representation for diagnosing Alzheimer's disease.
In addition to providing the spatial density of bone for deriving BMD, our decomposition learning can serve as representation learning for grading osteoporosis from a plain X-ray image.

\subsection{Image-to-image translation}
An image-to-image translation (I2I) model tries to map between source and target image domains. 
Typically, GANs were often used to train the I2I model \citep{isola_image--image_2017,zhu_unpaired_2017,wang_high-resolution_2018}.
We regard the decomposition task as an I2I task that aims to translate a plain X-ray image into the decomposed DRR.
Depending on the availability of data, the I2I model can be trained in a paired \citep{isola_image--image_2017} and unpaired \citep{zhu_unpaired_2017} manner while the paired-trained model usually learns faster and better.
Our method appreciated the paired training by utilizing 2D-3D registration \citep{otake_intraoperative_2012} to align X-ray images and decomposed DRRs, which are generated from 3D CT images.

\section{Method}
Fig. \ref{fig:method_overview} shows an overview of the proposed method that uses decomposition-based representation learning for BMD estimation. 
In decomposition training with the proposed BMD-GAN, an image synthesis model $G=\{M_E,M_D\}$, which consists of an encoder backbone $M_E$ and a decoder $M_D$, decomposes the X-ray image into the DRR of the proximal femur region [hereafter ``proximal femur region DRR (PF-DRR)''], whose average intensity provides the predicted BMD. 
We adopted HRFormer \citep{yuan_hrformer_2021} as the encoder backbone. 
Our BMD-GAN applies a hierarchical framework during the training in which the synthesis model is first trained to extract the pelvis and femur bones in stage-one and then the proximal femur region in stage-two. 
During regression (stage- three) training, the encoder backbone $M_E$ is inherited and connected to a Transformer-based regression head $M_R$ to become a regression network $R=\{M_E,M_R\}$ for end-to-end training of regressing BMD.
The regression head tries to regress BMD from a learned representation instead of the decomposed PF-DRR for efficient forwarding.
Note that our method is network-agnostic. Other backbone models and regression heads can be adopted to achieve decomposition learning and regression learning, respectively.

\subsection{Dataset construction}
% For better reproducibility, we describe our dataset construction method in detail. 
Fig. \ref{fig:dataset_const} illustrates the overall dataset construction procedure. 
In this study, we constructed two datasets for training different stages: 1) dataset A, which is used for stage-one training containing X-ray images; QCT; the 3D segmentation masks of the pelvis and femur, which were automatically obtained by applying the Bayesian U-net \citep{hiasa_automated_2020}; and GT bone DRR, which were created from QCT using 2D-3D registration \citep{otake_intraoperative_2012} followed by projection with the 3D mask, and 2) dataset B, which is used for other stages' training containing X-ray images; QCT; the 3D mask of the proximal femur region, which is obtained via manually labeled bony landmarks defined in \citep{uemura_development_2022} by an expert clinician; and the GT PF-DRR. 
The construction procedure of dataset B followed \citep{uemura_development_2022}. 
The intensity-based 2D-3D registration using gradient correlation similarity metrics and the CMA-ES optimizer \citep{otake_intraoperative_2012} was performed on each patient’s X-ray image and QCT. 
GT DXA-BMD and GT QCT-BMD values were associated with each patient's data in dataset B.
All X-ray images and DRRs were normalized into the 256 $\times$ 512 size via central cropping and resizing. 
The aspect ratio of the original X-ray images varied from  0.880 to 1.228 (width/height).
We first split them horizontally in half at the center.
Then the side with the target hip was reshaped to a predefined image size (256 $\times$ 512 in this experiment) by aligning the center of the image and cropping the region outside the image after resizing to fit the shorter edge of the width and height.

\begin{figure}[!t]
\centering
\includegraphics[width=\columnwidth]{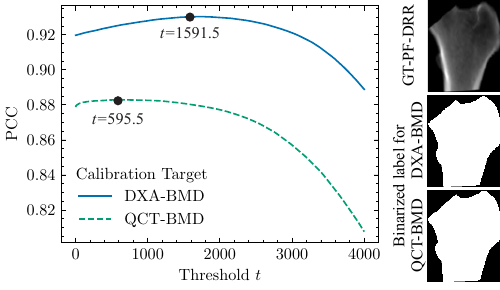}
\caption{
Tuning process of the threshold used to derive BMD from PF-DRR in stage two. GT-BMD is linearly fitted to the average intensity of the PF region in PF-DRR. PF-DRR is binarized by thresholding to determine the pixel count inside the PF region for calculating the average intensity. The line plot shows the threshold against PCC between GT-BMD and the BMD derived from PF-DRR, where the black points indicate the chosen threshold that maximizes PCC. The label images show the binarized results from a representative case using the chosen thresholds.
}
\label{fig:threshold_tuning}
\end{figure}

{\subsection{X-ray image Decomposition}
% % \DeclarePairedDelimiter{\norm}{\lVert}{\rVert} 
\newcommand{\image}[1]{\bm{\mathrm{#1}}}
\newcommand{\imgxray}{\image{I^{Xp}}}
\newcommand{\imgdrr}{\image{I^{DRR}}}
\newcommand{\Exraydrr}{\mathbbm E_{\left(\imgxray,\imgdrr\right)}}
\newcommand{\Exray}{\mathbbm E_{\imgxray}}
\newcommand{\imgxrayi}{\image{I}^{\image{Xp}}_i}
\newcommand{\imgdrri}{\image{I}^{\image{DRR}}_i}
The GAN with conditional discriminators was used to train the decomposition model. 
We followed most settings used in Pix2PixHD \citep{wang_high-resolution_2018}, including the multi-scale discriminators and the Feature Matching loss, among others. 
Instead of the ResNet Generator used in Pix2PixHD \citep{wang_high-resolution_2018} and CycleGAN \citep{zhu_unpaired_2017}, we adopted the state-of-the-art model HRFormer \citep{yuan_hrformer_2021}, a transformer-based model for segmentation, to be the backbone of the generator $G=\{M_E,M_D\}$, namely, the HRFormer Generator. 
Instead of using the hierarchical structure of the generator used in Pix2PixHD \citep{wang_high-resolution_2018}, we applied the HL framework in which a two-stage training is used. 
In stage one, the model is trained to decompose an X-ray image into pelvis and femur bones; in stage two, the target is transferred to the proximal femur region. 
During training, we used the adversarial loss $\mathcal{L}_{\mathrm{GAN}}$, which is defined as
\begin{equation}
\begin{split}
    \mathcal{L}_{\mathrm{GAN}}(G,D)
    & =\Exraydrr\bigl[\log D(\imgxray,\imgdrr)\bigr]\\
    & +\Exray\bigl[\log(1-D(\imgxray,G(\imgxray)))\bigr],
\end{split}
\label{eq:LGAN}
\end{equation}
where $D$, $\imgxray$, and $\imgdrr$ are the discriminator, X-ray image, and decomposed DRR, respectively.
We denote $\mathbbm{E}_{\bm{\mathrm{I^{Xp}}}}\triangleq\mathbbm{E}_{\bm{\mathrm{I^{Xp}}}\sim p_{\mathrm{data}}(\bm{\mathrm{I^{Xp}}})}$ and $\mathbbm{E}_{(\bm{\mathrm{I^{Xp}}},\bm{\mathrm{I^{DRR}}})}\triangleq\mathbbm{E}_{\bm{\mathrm{(I^{Xp}}},\bm{\mathrm{I^{DRR})}}\sim p_{\mathrm{data}}(\bm{\mathrm{I^{Xp}}},\bm{\mathrm{I^{DRR}}})}$. Furthermore, we used the Feature Matching loss $\mathcal{L}_{\mathrm{FM}}$ proposed in \citep{wang_high-resolution_2018}, given by
\begin{equation}
\begin{split}
    \mathcal{L}_{\mathrm{FM}}(G,D)=\Exraydrr\sum_{i=1}^{T}\frac{1}{N_{i}}\mathrm{FM}(G,D^{(i)}),
\end{split}
\end{equation}
where $D^{(i)}$, $T$, and $N_{i}$ denote the $i$th-layer feature extractor of the discriminator $D$, the total number of layers, and the number of elements in each layer, respectively. The layer feature matching criterion $\mathrm{FM}$ is defined as
\begin{equation}
\begin{split}
    \mathrm{FM}(G,D^{(i)})=\norm{D^{(i)}(\imgxray,\imgdrr)-D^{(i)}(\imgxray,G(\imgxray))}_1.
\end{split}
\end{equation}
We did not use a perceptual loss because a well-pretrained perceptual model is difficult to obtain using a limited-size dataset. We instead used a simple $\mathrm{L1}$ loss $\mathcal{L}_{\mathrm{L1}}$ defined as
\begin{equation}
\begin{split}
    \mathcal{L}_{\mathrm{L1}}(G)=\Exraydrr\norm{\imgdrr-G(\imgxray)}_1.
    \label{eq:Ll1}
\end{split}
\end{equation}
To maintain the consistency of the structure between the fake DRR $G(\imgxray)$ and the true DRR $\imgdrr$, we regularized the generator with the gradient matching constraints proposed in \citep{penney_comparison_1998}, using the gradient correlation loss \citep{hiasa_cross_2018} $\mathcal{L}_{\mathrm{GC}}$ defined as
\begin{equation}
    \mathcal{L}_{\mathrm{GC}}(G)=\Exraydrr GC(G).
\end{equation}
The criterion $\mathrm{GC}$ is defined as
\begin{equation}
\begin{split}
        \mathrm{GC}(G)
        & =\mathrm{NCC}(\nabla_{x}\imgdrr,\nabla_x G(\imgxray))\\
        & +\mathrm{NCC}(\nabla_{y}\imgdrr,\nabla_y G(\imgxray)),
\end{split}
\end{equation}
where $NCC(\bm{\mathrm{A}},\bm{\mathrm{B}})$ is the normalized cross-correlation of $\bm{\mathrm{A}}$ and $\bm{\mathrm{B}}$, and $\nabla_{x}$ and $\nabla_{y}$ are the $x$ and $y$ components of the gradient vector, respectively. We introduced hyper-parameters $\lambda_{L1}$, $\lambda_{GC}$, and $\lambda_{FM}$ to balance the importance of the loss terms summarized as
\begin{equation}
\begin{split}
    \mathcal{L}_{\mathrm{dec}}
    & =\lambda_{\mathrm{L1}}\mathcal{L}_{\mathrm{L1}}(G)+\lambda_{\mathrm{GC}}\mathcal{L}_{\mathrm{GC}}(G)\\
    & +\lambda_{\mathrm{FM}}\sum_{k=1,2,3}\mathcal{L}_{\mathrm{FM}}(G,D_k),
\end{split}
\end{equation}
where the multi-scale discriminators $D_1$, $D_2$, and $D_3$ were used under three resolutions as in \citep{wang_high-resolution_2018}. Thus, the full objective for decomposition training is defined as
\begin{equation}
    \min_G\biggl(\mathcal{L}_{\mathrm{dec}}+\max_{D_1,D_2,D_3}\Bigl(\sum_{k=1,2,3}\mathcal{L}_{\mathrm{GAN}}(G,D_k)\Bigr)\biggr).
\end{equation}
Both stage one and stage two training use the same loss functions.

Once the decomposition was completed, the average intensity of the predicted PF-DRR was calculated. 
Note that pixels whose intensities were equal to or greater than the threshold $t$ were averaged for deriving BMD.
The PF-DRR-average of all training datasets was linearly fitted to DXA-BMD and QCT-BMD to obtain the slope and intercept, which were used to convert the PF-DRR-average to the BMDs of the test dataset.
The threshold $t$ tuning procedure with GT-PF-DRR is shown in Fig. \ref{fig:threshold_tuning}, where we empirically chose values for different situations.

\subsection{Regression from learned representation}
For the regression head $M_R$, we used several Transformer blocks before the final fully-connected layer to extract BMD-related features from a representation learned by the encoder backbone from an X-ray image.
We borrowed the architecture of the CoAtNet \citep{dai_coatnet_2021} as our regression head for its high performance on image classification tasks by combining convolution blocks and Transformer blocks. 
However, we only used the Transformer blocks from it, which applies input-independent relative attention. 
Unlike in CoAtNet whose first Transformer block takes a representation (i.e., feature maps) with $\frac{1}{8}$ size of the input image, our regression head takes a higher-resolution representation with $\frac{1}{4}$ size learned by the encoder $M_E$. 
The inherited encoder and the regression head form our regression network $R$.

During training, a simple L1 loss was used.
In addition, we introduced a predefined sample weight into the loss to improve unbalanced learning since BMD values are not uniformly distributed in the real world.
The sample weight is calculated according to the distance $d$ between the sample's BMD $y$ and the average BMD $\hat{y}$ of the dataset.
The distance is defined as
\begin{equation}
    d=|y-\hat{y}|.
\end{equation}
For each X-ray image, its sample weight $w$ is defined as
\begin{equation}
    w=1.5-\frac{d-d_{min}}{d_{max}-d_{min}},
\end{equation}
where the $d_{min}$ and $d_{max}$ are the minimum and maximum distances in the dataset.
Thus, the objective of regression training is defined as
\begin{equation}
    \min_{R}\mathbbm E_{(\imgxray,y,w)}w\bigl|y-R(\imgxray)\bigr|,
\end{equation}
where we denote $\mathbbm{E}_{(\imgxray,y,w)}\triangleq\mathbbm{E}_{(\imgxray,y,w)\sim p_{\mathrm{data}}(\imgxray,y,w)}$.
Because we treat the DXA-BMD and QCT-BMD estimations as independent tasks, the sample weight for an X-ray image can be different when regressing different BMDs.
}
\begin{table}[!t]
\caption{Datasets Cohort}
\centering
\begin{adjustbox}{width=\columnwidth}
\begin{tabular}{lll}
Item                                    & Dataset A    & Dataset B \\ \hline
Number of X-ray images                  & 525          & 1204  \\
Number of patients                      & 275          & 305   \\
Female, n (\%)                          & 236 (85.8)   & 250 (82.0)    \\
Mean age (sd), years                    & 58.6 (14.3)  & 57.6 (14.9)   \\
Mean BMI (sd), $kg/m^2$                 & 23.1 (3.8)          & 23.5 (4.0)  \\
Mean DXA-BMD (sd), $g/cm^2$             & n/a          & 0.759 (0.151)   \\
Median T-score (IQR)                    & n/a          & -1.2 (-2.1, 0.4)   \\
Osteoporosis w.r.t. DXA-BMD, n (\%)     & n/a          & 50 (16.4)   \\
Mean CT-BMD (sd), $g/cm^3$              & n/a          & 290 (57)  
\end{tabular}
\end{adjustbox}
\label{table:dataset}
\end{table}

\section{Experiments and Results}
We validated our method under the constraint of limited datasets.
Based on the constructed datasets (A and B), we conducted two experiments for 1) evaluating the performance of the BMD estimation of our method and 2) conducting an ablation study on the proposed hierarchical learning under different backbone models.

\subsection{Dataset}
\begin{figure}[!t]
\centering
\includegraphics[width=\columnwidth]{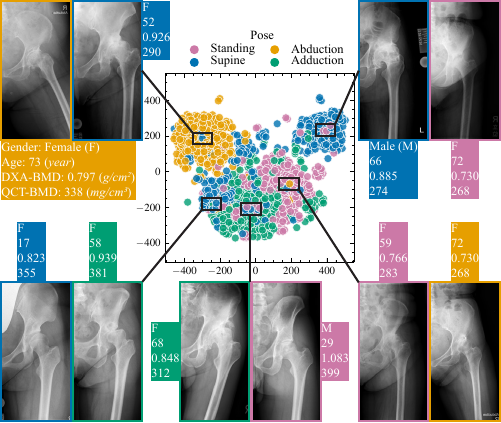}
\caption{
T-SNE visualization of X-ray images in dataset B with data collected from 305 patients. Each data point represents an X-ray image. The horizontal and vertical axes are the first and second dimensions in T-SNE space, respectively. The X-ray images were collected with different poses (standing, adduction, abduction, and supine), suggesting a high variability in the dataset and a high potential for various validations.}
\label{fig:dataset_tsne}
\end{figure}

The two datasets (A and B) used in the experiments are summarized in Table \ref{table:dataset}.
Ethical approval was obtained from the Institutional Review Boards (IRBs) of the institutions participating in this study (IRB approval numbers: 21115 for Osaka University Hospital and 2021-M-11 for the Nara Institute of Science and Technology). 
The constructed dataset A contained 275 cases. 
Each case had an X-ray image, and its paired bone DRRs of the left and right sides were split by the vertical middle line, resulting in 525 image pairs after excluding the images with hip implants. 
The constructed dataset B contained 305 cases obtained retrospectively from 305 patients (250 females) who underwent primary total hip arthroplasty between May 2011 and December 2015.
Each case has three or four X-ray images with different poses and their paired PF-DRR of one side with their ground-truth DXA-BMD and QCT-BMD. 
Fig. \ref{fig:dataset_tsne} visualized the T-SNE of the collected multi-pose X-ray images of dataset B, which includes poses of standing, adduction, abduction, and supine, suggesting high potential for conducting multiple clinical validations.
The calibration phantom (B-MAS200, Kyoto Kagaku, Kyoto, Japan) \citep{uemura_automated_2021}, which is used to convert radiodensity [in Hounsfield units] to the ground truth QCT-BMD (in $\mathrm{mg/cm^3}$), contains known densities of hydroxyapatite $\mathrm{Ca_{10}(PO_{4})_{6}(OH)_{2}}$. 
All CT images used in this study were obtained using the OptimaCT660 scanner (GE Healthcare Japan, Tokyo, Japan), and all DXA images of the proximal femur were acquired for the operative side (Discovery A, Hologic Japan, Tokyo, Japan) to obtain the ground truth DXA-BMD.

\begin{figure*}[!t]
\centering
\includegraphics[width=\textwidth]{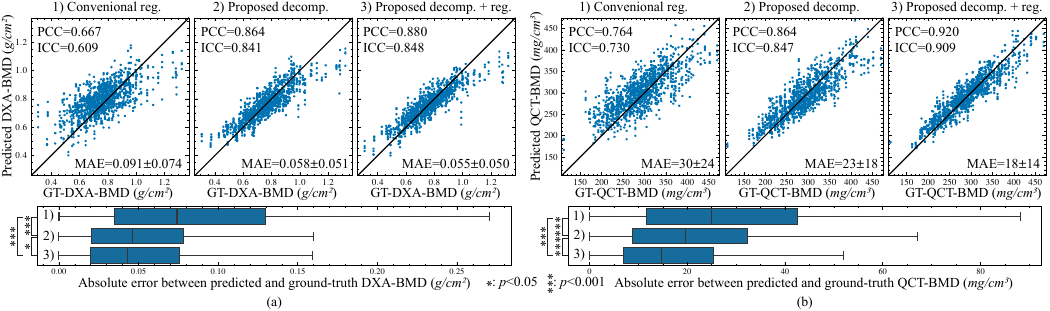}
\caption{Correlation and absolute error between predicted and GT BMDs. (a) Prediction of DXA-BMD. (b) Prediction of QCT-BMD. 1) Conventional reg. is the direct regression method trained and tested using the pre-cropped ROI of the proximal femur of X-ray images. 2) Proposed decomp. is the proposed decomposition method that predicts 2D BMD distribution. 3) Proposed decomp. + reg. is the proposed regression method based on method 2).}
\label{fig:315_bmd_result}
\end{figure*}

\begin{table*}[!t]
\caption{Summary of BMD estimation experiment result}
\centering
\small
\begin{adjustbox}{width=.95\textwidth}
\begin{tabular}{c|cl|ccccc|ccccc}
\multirow{2}{*}{\begin{tabular}[c]{@{}c@{}}Pose in\\ training dataset\end{tabular}}                   & \multicolumn{2}{c|}{\multirow{2}{*}{Method}}                                & \multicolumn{5}{c|}{DXA-BMD Estimation Accuracy}                                                  & \multicolumn{5}{c}{QCT-BMD Estimation Accuracy}                                                \\ \cline{4-13} 
                                                                                                      & \multicolumn{2}{c|}{}                                                       & PCC$\uparrow$  & ICC$\uparrow$  & MAE$\downarrow$& SEE$\downarrow$& RMS-CV$\downarrow$        & PCC$\uparrow$  & ICC$\uparrow$  & MAE$\downarrow$& SEE$\downarrow$& RMS-CV$\downarrow$     \\ \hline
\multirow{9}{*}{\begin{tabular}[c]{@{}c@{}}Standing,\\ Supine,\\ Abduction,\\ Adduction\end{tabular}} & \multicolumn{2}{c|}{Conventional Regression}                                & 0.667          & 0.609          & 0.091          & 0.117          & 6.02\%                 & 0.764          & 0.730          & 30.3           & 38.6           & 5.99\%              \\ \cline{2-3}
                                                                                                      & \multicolumn{1}{c|}{\multirow{6}{*}{Proposed}} & Decomposition using QCT    & 0.864          & {\ul 0.841}    & 0.058          & 0.077          & \textbf{3.27\%}        & 0.861          & 0.847          & 23.1           & 29.3           & 3.79\%              \\
                                                                                                      & \multicolumn{1}{c|}{}                          & +Regression (freeze $M_E$) & {\ul 0.874}    & 0.837          & 0.057          & {\ul 0.076}    & 3.40\%             & 0.914          & 0.904          & 18.4           & 23.6               & 3.61\%          \\
                                                                                                      & \multicolumn{1}{c|}{}                          & +Regression (tune $R$)     & \textbf{0.880} & \textbf{0.848} & \textbf{0.055} & \textbf{0.074} & {\ul 3.35\%}             & {\ul 0.920}    & {\ul 0.909}    & {\ul 17.9}     & {\ul 22.9}     & \textbf{3.36\%}     \\ \cline{3-3}
                                                                                                      & \multicolumn{1}{c|}{}                          & Decomposition using CT     & 0.836          & 0.814          & 0.062          & 0.083          & 3.53\%                 & 0.871          & 0.852          & 22.2           & 28.4           & 3.78\%              \\
                                                                                                      & \multicolumn{1}{c|}{}                          & +Regression (freeze $M_E$) & 0.864          & 0.829          & 0.057          & 0.078          & 3.54\%                 & 0.915          & 0.905          & 18.1           & 23.4           & 3.73\%             \\
                                                                                                      & \multicolumn{1}{c|}{}                          & +Regression (tune R)       & 0.872          & 0.838          & 0.056          & 0.076          & 3.46\%                 & \textbf{0.922} & \textbf{0.910} & \textbf{17.6}  & \textbf{22.7}  & {\ul 3.54\%}          \\ \cline{2-13} 
                                                                                                      & \multicolumn{2}{c|}{GT-PF-DRR from QCT}                                     & 0.930          & 0.928          & 0.040          & 0.055          & 1.39\%                 & 0.883          & 0.877          & 20.8           & 27.0           & 1.24\%              \\
                                                                                                      & \multicolumn{2}{c|}{GT-PF-DRR from CT}                                      & 0.921          & 0.918          & 0.044          & 0.059          & 1.49\%                 & 0.893          & 0.888          & 19.9           & 25.8           & 1.36\%              \\ \hline
\multirow{6}{*}{\begin{tabular}[c]{@{}c@{}}Standing,\\ Supine\end{tabular}}                           & \multicolumn{1}{c|}{\multirow{6}{*}{Proposed}} & Decomposition using QCT    & 0.855          & {\ul 0.837}    & 0.059          & 0.079          & {\ul 3.72\%}    & 0.851          & 0.833          & 23.8           & 30.3                    & \textbf{3.77\%} \\
                                                                                                      & \multicolumn{1}{c|}{}                          & +Regression (freeze $M_E$) & {\ul 0.862}    & 0.835          & {\ul 0.058}    & {\ul 0.078}    & 4.02\%                 & {\ul 0.909}    & {\ul 0.897}    & {\ul 18.9}     & {\ul 24.3}     & 3.99\%              \\
                                                                                                      & \multicolumn{1}{c|}{}                          & +Regression (tune $R$)     & \textbf{0.866} & \textbf{0.840} & \textbf{0.057} & \textbf{0.076} & 3.93\%                 & \textbf{0.913} & \textbf{0.900} & \textbf{18.6}  & \textbf{23.8}  & {\ul 3.85\%} \\ \cline{3-3}
                                                                                                      & \multicolumn{1}{c|}{}                          & Decomposition using CT     & 0.833          & 0.803          & 0.063          & 0.084          & \textbf{3.60\%}    & 0.848          & 0.832          & 24.0           & 30.5               & 4.49\%              \\
                                                                                                      & \multicolumn{1}{c|}{}                          & +Regression (freeze $M_E$) & 0.833          & 0.808          & 0.063          & 0.084          & 4.18\%                 & 0.898          & 0.888          & 19.9           & 25.7           & 4.19\%              \\
                                                                                                      & \multicolumn{1}{c|}{}                          & +Regression (tune $R$)     & 0.847          & 0.823          & 0.061          & 0.081          & 4.28\%                 & 0.903          & 0.893          & 19.4           & 25.0           & 4.12\%         
\end{tabular}
\end{adjustbox}
\label{tab:summary_bmd_est}
\end{table*}

\subsection{BMD estimation experiment}
% Experimental setting
\subsubsection{Experimental settings}
In this experiment, we tried to estimate DXA-BMD and QCT-BMD as independent tasks.
Dataset A was used for training stage one of the proposed method, containing 492 pairs of X-ray images and GT bone DRR.
Five-fold cross-validation was performed on stages two and three using dataset B.
We compared our method with the conventional one proposed by \citet{hsieh_automated_2021}, which directly regresses BMD from an ROI of an X-ray image obtained using a localization model that tries to find the proximal femur. 
When applying the conventional method, we used pre-cropped ROI of the proximal femur of X-ray images to simulate the most ideal situation.
The proposed and conventional methods were trained using the same number of patient data to evaluate the effects of utilizing CT data in training by the proposed method.
We evaluated BMD estimation performance by Pearson correlation coefficient (PCC), intraclass correlation coefficient (ICC), mean absolute error (MAE), and standard error of estimate (SEE).
Statistical significance was evaluated using the single-factor repeated measures analysis of the variance model. 
P-values were used to denote statistical significance.
We also evaluated the reproducibility under the four poses (standing, supine, abduction, and adduction) using root mean square of the coefficient of variation (RMS-CV).
The results will be reported in Sec. \ref{sec:overall_performance}.
In addition to using QCT and X-ray images of four poses, we also conducted three more experiments by 1) training with the limited-pose dataset, 2) training with uncalibrated CT, and 3) testing with compressed X-ray images for robustness validation against 1) pose variation, 2) un-calibration, and 3) image compression reported in Sec. \ref{sec:exp_pose}, Sec. \ref{sec:exp_ct}, and Sec. \ref{sec:exp_compression}, respectively.
These additional experiments were designed to validate the generalizability of the proposed method considering the situations that are more likely to happen in real-world applications.
For the training of all stages, we used the learning rate policy of stochastic gradient descent with warm restarts \citep{loshchilov_sgdr_2017} and AdamW optimizer \citep{loshchilov_decoupled_2019}.
In stage three training, we fixed the weights of encoder $M_E$ by several epochs at the beginning and then trained $R$ by the remaining epochs.

% For comparison between 4-pose and 2-pose.
\begin{figure*}[!t]
\centering
\includegraphics[width=\textwidth]{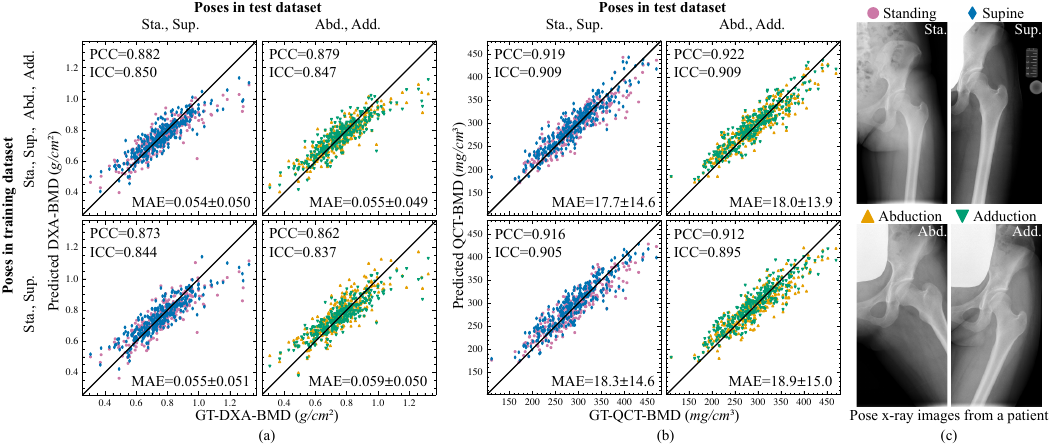}
\caption{Pose-wise correlations between predicted and ground-truth BMD by the proposed method in experiments of training with (upper row) four poses and (lower row) two poses. (a) Prediction of DXA-BMD. (b) Prediction of QCT-BMD. (c) Typical X-ray image of each pose.}
\label{fig:315_pose_bmd_result}
\end{figure*}

\begin{figure}[!t]
\centering
\includegraphics[width=\columnwidth]{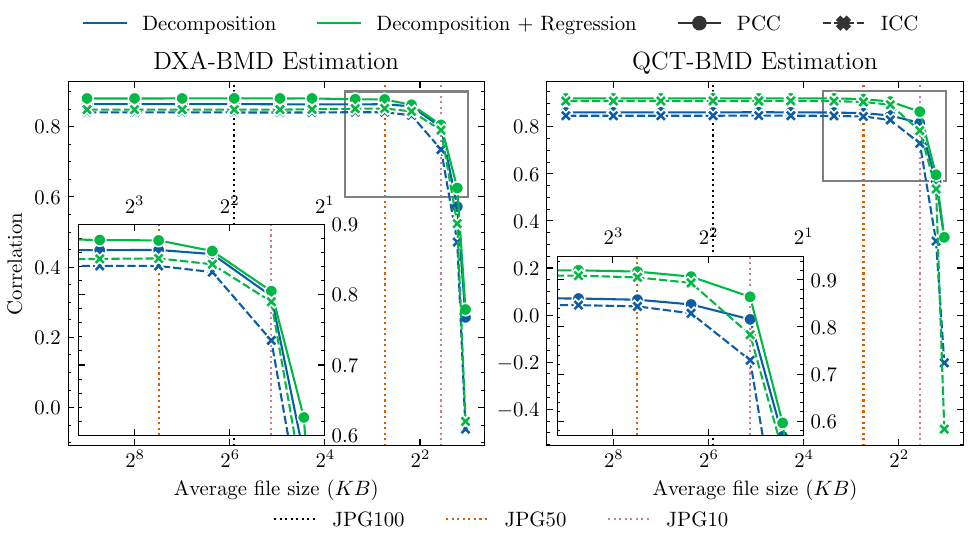}
\caption{Line plots showing the BMD estimation performance by our method affected by the compression of X-ray images with fewer bits and the JPEG compression algorithm when targeting DXA-BMD (left) and QCT-BMD (right). 
The horizontal axes indicate the average file sizes of images after compression under the size of 512 $\times$ 256. The 
 vertical axes show the corresponding correlation scores obtained by our method.}
\label{fig:compress_plots}
\end{figure}
% Results
\subsubsection{Overall performance of BMD estimation}
\label{sec:overall_performance}
Comparisons of BMD estimation performance between methods are shown in Fig. \ref{fig:315_bmd_result}. 
The proposed decomposition method (``proposed decomp.'' in Fig. \ref{fig:315_bmd_result}), which is the $G$ model trained with hierarchical learning, improved DXA-BMD estimation PCC and ICC from 0.667 and 0.609 to 0.864 and 0.841, respectively. 
A similar trend was also observed for the QCT-BMD estimation task, where the PCC and ICC were improved from 0.764 and 0.730 to 0.861 and 0.841 by the proposed decomposition method, respectively. 
The estimation performance was further improved by learning regression from a representation learned by encoder $M_E$ via decomposition training (``proposed decomp. + reg.'' in Fig. \ref{fig:315_bmd_result}), achieving the highest PCC and ICC of 0.880 and 0.848s, respectively, for DXA-BMD estimation, and 0.920 and 0.909, respectively, for QCT-BMD estimation. 
An evaluation summary for full experiments and metrics is shown in Table \ref{tab:summary_bmd_est}.
Under the guidance of decomposition training, regressing BMD already achieved high scores while freezing model weights of encoder $M_E$, demonstrating successful representation learning by decomposition.
Furthermore, final tuning for the encoder and regression head improved performance, achieving the highest scores on all metrics we used.
Notice that the decomposition-guided regression even outperformed the derivation of QCT-BMD from GT-DRR on most metrics.
RMS-CV values for our methods were 3.27 to 3.79\%, while the acceptable RMS-CV in the DXA-BMD measurement is 1.8\% \citep{lewiecki_best_2016}.
However, this criterion assumes that the patient pose should be reproduced as much as possible (whereas in our conditions, the patient poses vary considerably). 
A video showing prediction samples can be found in the supplemental materials.

\subsubsection{Robustness against pose variation}
\label{sec:exp_pose}
Unlike DXA scanning, which requires patients to be in a supine pose, X-ray imaging does not restrict the scanning pose, resulting in pose variations, which imply challenges when predicting the BMD from different poses.
To validate robustness against pose variation, we conducted additional experiments using training data with poses limited to standing and supine, which are more commonly accessible in routine clinical practice, reducing the number of training data from 963 to 481 per fold.

Fig. \ref{fig:315_pose_bmd_result} compares the BMD estimation performance of the proposed method between training with full poses (standing, supine, abduction, adduction) and limited poses (standing, supine).
For the seen poses (standing and supine), limited-pose training showed a small reduction of PCCs from 0.882 to 0.873 and from 0.919 to 0.916 for DXA-BMD and QCT-BMD, respectively.
For the unseen poses, abduction and adduction, limited-pose training showed a reduction of PCCs from 0.879 to 0.862 and from 0.922 to 0.912 for DXA-BMD and QCT-BMD, respectively.
Although such slight degradations were observed, statistical tests via the Tukey HSD pose-wise showed $p$-values all higher than 0.05, failing to indicate significant differences between training with full and limited-pose data.
Furthermore, we checked the prediction consistency between poses.
The proposed method showed all PCCs higher than 0.9 for the densely compared poses on not only seen poses but also unseen poses.
Detailed pose prediction consistency can be found in the supplementary materials.

\subsubsection{Robustness against image compression}
\label{sec:exp_compression}
We validated the proposed method for robustness against image compression since the compressed images are much easier to obtain in routine clinical practice.
This compression validation is based on full-pose training experiments reported in Sec \ref{sec:overall_performance}.
We evaluated the performance of the trained models using X-ray images compressed by the JPEG lossy method.

Fig. \ref{fig:compress_plots} shows the relations between the average file sizes of compressed images and the performance of BMD estimation.
As the file size has been significantly reduced by lower precision and the JPEG algorithm with 100 quality, the proposed decomposition-based regression method's performance remains almost unchanged with those with no compression.
Furthermore, we explored the tolerance limitation by inputting super-compressed images after reducing the JPEG quality.
We found that using the half quality, JPEG 50, slightly hurt overall performance, reducing PCCs from 0.880 and 0.920 to 0.877 and 0.918 for DXA-BMD and QCT-BMD estimations, respectively, with only 1.3\% average file size of the original, demonstrating high robustness against significant loss of available information in X-ray images.
Severe performance degradation was observed from JPEG 10, where the PCCs were reduced to 0.790 and 0.783 for DXA-BMD and QCT-BMD, respectively.
Visualizations of the compression impact on our method can be found in the supplementary materials.

% Uncalibrated CT Results
\subsubsection{Robustness uncalibrated CT}
\label{sec:exp_ct}
Our method utilized CT information to improve accuracy and efficiency; however, quantitative CT with phantom calibration is required, while large-scale CT datasets, which do not contain phantoms for calibration, were not utilized.
On the other hand, uncalibrated CT introduced the domain-shifting problem, which is a long-standing challenge in image processing.
We validated the proposed method by using uncalibrated CT.

Without using QCT, the proposed method of Decomp. + Reg. achieved a PCC of 0.872 and 0.922 and ICC of 0.838 and 0.910 for the DXA-BMD and QCT-BMD estimation tasks with full-pose training, respectively.
For limited-pose training, the proposed method achieved a PCC of 0.847 and 0.903 and ICC of 0.823 and 0.893 for the DXA-BMD and QCT-BMD estimation tasks, respectively.
The Tukey HSD test was performed on a pose-by-pose basis on the predictions given by the proposed method with QCT and uncalibrated CT for full-pose training and limited-pose training separately, and no significant differences were found between them.
The results were expected as the CT data obtained in our dataset were all from the same type of scanner.
However, the results could be considered as an upper bound of what can be achieved by training using uncalibrated CT.

\subsubsection{Source for large error}
\begin{figure}[!t]
\centering
\includegraphics[width=\columnwidth]{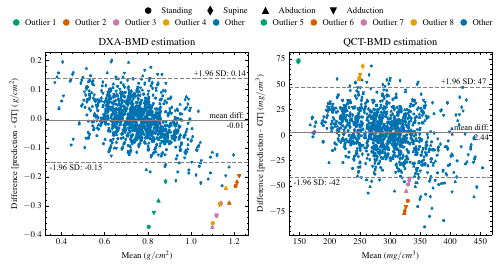}
\caption{Bland-Altman plots for the DXA-BMD estimation task (left) and the QCT-BMD estimation task (right).
Case-level outliers are indicated for whose points of all poses are either above +1.96SD or below -1.96SD.}
\label{fig:315_bland_altman}
\end{figure}

We investigated the error source of the proposed method in the experiments that use full-pose training data and QCT-derived DRR.
Fig. \ref{fig:315_bland_altman} shows Bland-Altman plots of the DXA-BMD estimation task (left) and the QCT-BMD estimation task (right), respectively, by the proposed decomposition-guided regression method.
Points beyond the 95\% limits of agreement (i.e., mean $\pm$ 1.96SD) are considered sample outliers, where the SD is the standard deviation of the difference between prediction and ground truth. 
Each point represents an X-ray image in the corresponding task.
The trend showed a bias underestimating and overestimating high-BMD and low-BMD patients, respectively.
One possible bias source could be the lack of training data in those BMD areas, which introduced unbalanced learning, which is a long-standing problem in real-world applications \citep{li_targeted_2022,alshammari_long_2022}.
Bland-Altman plots by the conventional and proposed decomposition methods can be found in supplemental materials.

\begin{figure}[!t]
\centering
\includegraphics[width=\columnwidth]{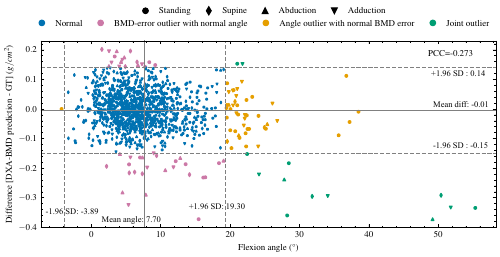}
\caption{Patients' hip-joint (flexion) angle against BMD estimation error. Outliers concerning BMD error and angle are indicated using mean $\pm$ 1.96SD limits of agreement.}
\label{fig:315_flexsion_vs_error}
\end{figure}

\begin{figure*}[!t]
\centering
\includegraphics[width=\textwidth]{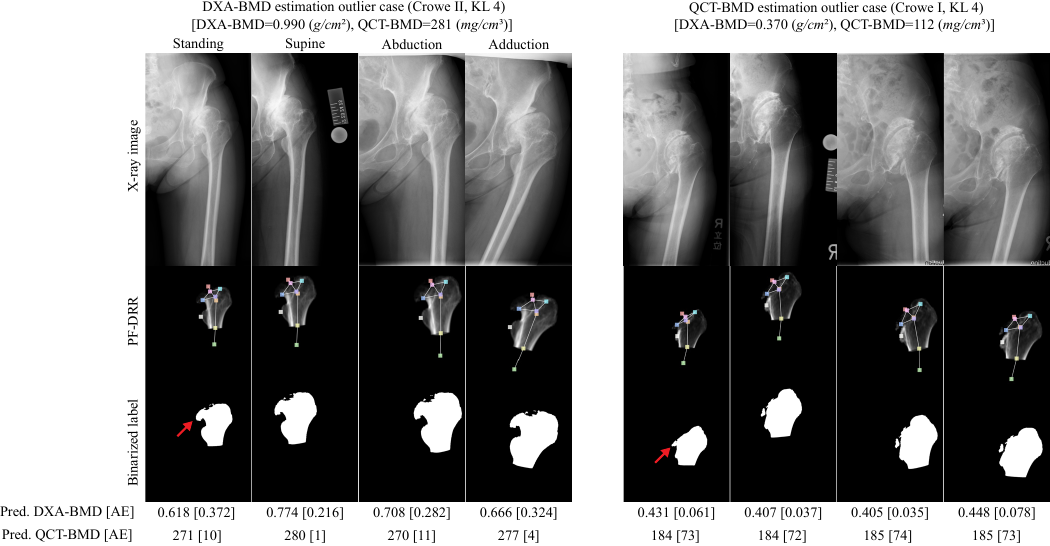}
\caption{Outlier cases in the DXA-BMD estimation task (left) and the QCT-BMD estimation task (right).
The PF-DRR and label images are cropped with the same bounding box to align the horizontal view.
The red arrows indicate the edge of the femoral head included in the PF region due to hip deformation, potentially introducing errors in BMD estimation.}
\label{fig:315_outlier_cases}
\end{figure*}
For the DXA-BMD estimation task, there are 57 outlier samples in Fig. \ref{fig:315_bland_altman}. 
We carefully checked their X-ray imaging conditions, including the tension peak of the X-ray tube in kV, exposure, and magnification; however, we found no particular trend or strong correlation with the estimation error.
Additionally, we checked the correlation between the estimation error and the patient's femur position with respect to the pelvis (i.e., joint angle). Fig. \ref{fig:315_flexsion_vs_error} shows the correlation to the flexion angle, where the estimation outlier rates are 3.93\% and 22.41\% in the normal- and abnormal-degree patients, respectively, implying potential estimation error introduced by patient positioning \citep{uemura_effect_2023}. When correlating to other angles, the estimation outlier rates in the normal- and abnormal-degree patients are 4.18\% and 11.11\%, respectively, for the abduction angle, and 5.37\% and 8.96\%, respectively, for the external rotation (ER) angle.
Scatter plots representing the correlation between the estimation error and abduction and ER angles can be found in supplemental materials.
Another possible error source could be hip diseases in patients.
In our dataset, most patients had developed hip diseases (e.g., osteonecrosis and osteoarthritis), and some even had developed severe forms of these diseases, which may have caused the measurement to be inaccurate.
We identified four outlier patients\ (outliers 1 to 4) in Fig. \ref{fig:315_bland_altman} (left), whose points of all poses were indicated as outlier samples.
The DXA-BMD of the four outlier patients were in the high-BMD area.
Fig. \ref{fig:315_outlier_cases} (left) shows the X-ray images and their paired PF-DRRs of the outlier-1 patient.
The patient's Kellgren-Lawrence grade \citep{kohn_classifications_2016} was 4 (the highest level), and the patient was classified as level II of Crowe \citep{sugano_morphology_1998}; i.e., high-severity hip osteoarthritis.
Referring to the QCT-BMD of outlier 1, we found that the QCT-BMD is below the distribution average, and our method accurately predicted it.

A similar analysis was performed for the QCT-BMD estimation task as well.
Fig. \ref{fig:315_bland_altman} (right) indicated 18, 17, 17, and 16 outlier samples in the standing, supine, abduction, and adduction poses, respectively.
Fig. \ref{fig:315_outlier_cases} (right) shows visualization for outlier-5 patients.
The PF of this patient had almost no neck part, where the clear clinical definition for the landmarks used to measure QCT-BMD is not available for such cases, resulting in undefined measurements.
Sources of error other than diseases and training data scales are open for deeper exploration.

\subsubsection{Implementation details}
The encoder backbone we adopted is HRFormer-B, which is a middle-size model.
Our decoder followed the upsampling parts of the GlobalGenerator in Pix2PixHD, consisting of several convolution layers and transpose-convolution layers.
In the regression head, we used Transformer blocks of S3-$\mathrm{TFM_{Rel}}$ and S4-$\mathrm{TFM_{Rel}}$ from CoAtNet
We replaced the Batch Normalization and Instance Normalization with Group Normalization, except for the Layer Normalization used in Transformer.
We set $\lambda_{L1}$, $\lambda_{GC}$, and $\lambda_{FM}$ to 100, 1, and 10, respectively.
The initial learning rate and weight decay were $6\times10^{-5}$ and $1\times10^{-2}$, respectively, except for tuning the end-to-end regression in stage three, which used $8\times10^{-6}$ as the initial learning rate.
For all our methods (i.e., decomposition and decomposition + regression), we train 630 epochs in stage one.
For the decomposition method, we trained 150 epochs in stage two.
For decomposition + regression, we trained 630 and 310 epochs in stages two and three, respectively, and tuned end-to-end regression with 70 epochs.
Data augmentation included random rotation($\pm$25), shear($\pm$9), translation($\pm$0.3), scaling($\pm$0.3), brightness($\pm$0.5), contrast($\pm$0.5), and horizontal and vertical flipping.
When implementing the conventional method for BMD estimation, we followed most settings and training protocols; however, the total number of epochs was set to 400, and we did not perform validation to capture the best model during training because we observed worse performance with our limited-size dataset.

\begin{figure*}[!t]
\centering
\includegraphics[width=.95\textwidth]{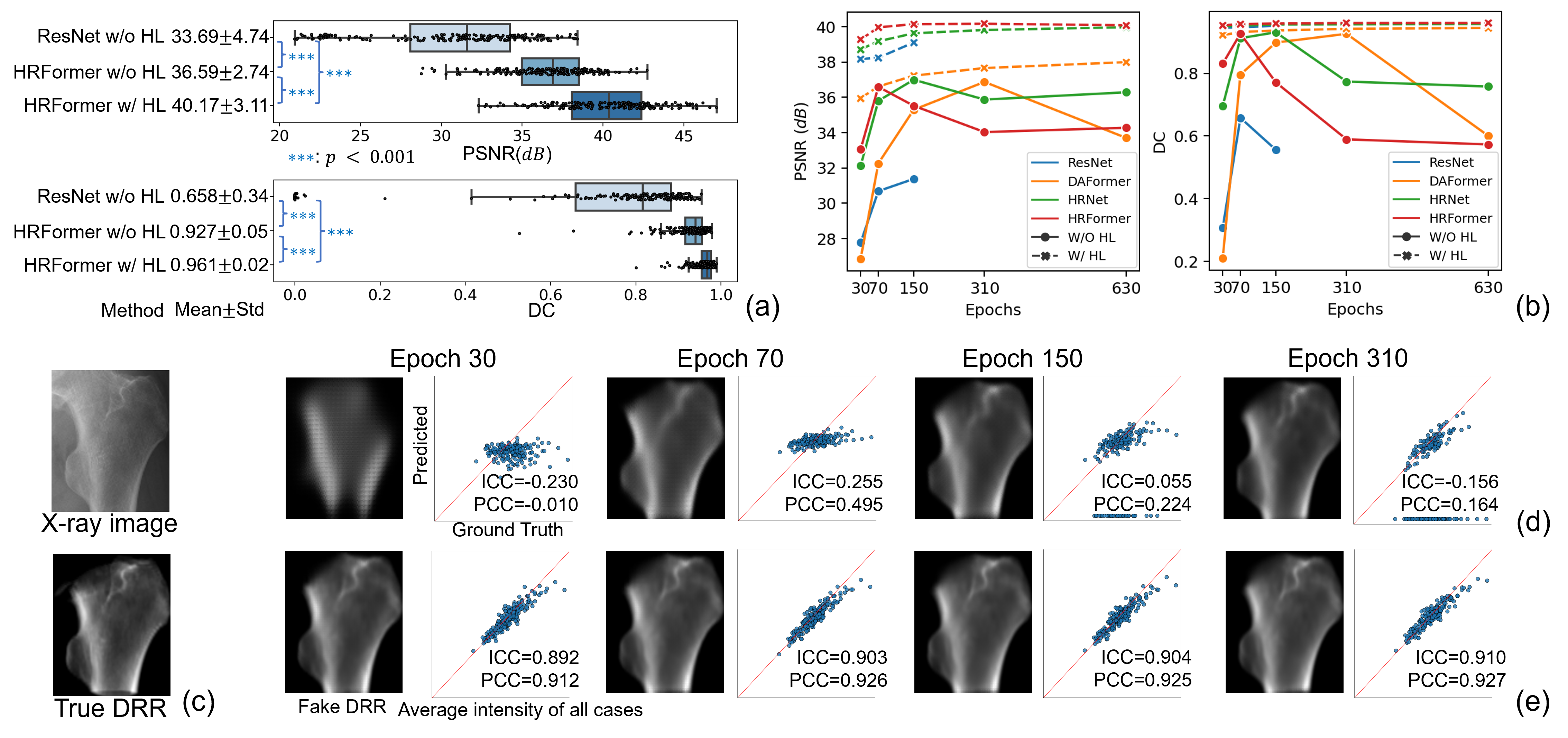}
\caption{Ablation study results of X-ray image decomposition.
(a) Evaluation of image decomposition for ResNet (baseline) without HL, HRFormer without HL, and HRFormer with HL.
(b) The Convergence analysis of each backbone without and with HL. (c) The ROI of the input X-ray image and GT-DRR, and the comparison of the training progress between (d) without HL and (e) with HL.
Without HL in (d), predicted BMD became zero at 150 and 300 due to instability of decomposition in several cases}
\label{fig:decomp_perform}
\end{figure*}

\subsection{Ablation study on hierarchical learning}
\label{sec:exp_ablation}
% MICCAI paper content
\subsubsection{Experimental settings}The ablation study aims to compare decomposition performances between 1) model backbones and 2) models with and without hierarchical learning. 
In addition to the ResNet and HRFormer, we compared two more models that were used for semantic segmentation as the backbone of the generator--DAFormer \cite{hoyer_daformer_2022}, and HRNetV2 \cite{wang_deep_2021}, namely, DAFormer Generator and HRNet Generator, respectively.
Though these state-of-the-art models have been proven to have high performance in segmentation tasks, their ability to decompose images has not been thoroughly assessed.
We set the ResNet Generator without HL as the baseline method.
Dataset A was used to train stage-one models.
For stage two, we only used 200 cases of standing pose from dataset B with five-fold cross-validation.
We also compared our best decomposition method with the conventional method \cite{hsieh_automated_2021} on the DXA-BMD estimation task using our limited dataset.
To evaluate the performance of image decomposition, we used the peak signal-to-noise ratio (PSNR), multi-threshold dice coefficient (DC), ICC, and PCC of the average intensity of PF-DRR.
To evaluate BMD estimation, we used ICC, PCC, MAE, and SEE.

\begin{table*}[!t]
\caption{Summary of the ablation study results.}
\centering
\small
\begin{tabular}{|l|cccc|ccccc|}
\hline
\multicolumn{1}{|c|}{}                         & \multicolumn{4}{c|}{Image Decomposition Accuracy}                                  & \multicolumn{5}{c|}{DXA-BMD Estimation Accuracy} \\ \cline{2-10} 
\multicolumn{1}{|c|}{\multirow{-2}{*}{Method}} & Mean PSNR              & Mean DC             & ICC & PCC & Mean AE             & SEE & ICC & PCC & PCC w.r.t. QCT-BMD                   \\ \hline
ResNet                                         & 30.688            & 0.658          & -0.278                & 0.006                 & 0.117          & 0.157                 & -0.024                & -0.208                & -0.130                    \\
+ HL                                           & \textbf{39.105}   & \textbf{0.952} & \textbf{0.866}        & \textbf{0.894}        & \textbf{0.057} & \textbf{0.074}        & \textbf{0.872}        & \textbf{0.879}        & \textbf{0.818}            \\ \hline
DAFormer                                       & 36.874            & 0.926          & 0.538                 & 0.671                 & 0.085          & 0.114                 & 0.639                 & 0.680                 & 0.632                      \\
+ HL                                           & \textbf{37.994}   & \textbf{0.945} & \textbf{0.853}        & \textbf{0.875}        & \textbf{0.057} & \textbf{0.078}        & \textbf{0.856}        & \textbf{0.865}        & \textbf{0.799}             \\ \hline
HRNet                                          & 36.996            & 0.931          & 0.369                 & 0.650                 & 0.097          & 0.127                 & 0.537                 & 0.581                 & 0.640                      \\
+HL                                            & \textbf{39.971}   & \textbf{0.958} & \textbf{0.883}        & \textbf{0.920}        & \textbf{0.057} & \textbf{0.074}        & \textbf{0.870}        & \textbf{0.878}        & \textbf{0.843}             \\ \hline
HRFormer                                       & 36.594            & 0.927          & 0.255                 & 0.495                 & 0.109          & 0.143                 & 0.313                 & 0.400                 & 0.498                      \\
+ HL                                           & \textbf{40.168}   & \textbf{0.961} & \textbf{0.910}        & \textbf{0.927}        &\textbf{0.053}  & \textbf{0.071}        & \textbf{0.882}        & \textbf{0.888}        & \textbf{0.853}             \\ \hline
\end{tabular}
%\caption{Summary of the experimental results.}
\label{tab:ablation}
\end{table*}
\subsubsection{Decomposition performance} The decomposition accuracy, as evaluated by PSNR and DC is shown in Fig. \ref{fig:decomp_perform} (a), where a significant improvement in HL was observed.
The high performance of the HRFormer Generator with HL in DC indicated the ability to maintain the silhouette of the decomposed structure. The high PSNR suggested the superior capability of the quantitative decomposition compared with the same generator without HL and the baseline method. 
Fig. \ref{fig:decomp_perform} (b) shows the training progress for each backbone with and without HL, in which the robust convergence was achieved consistently using HL, even with only a few epochs. 
One case was randomly chosen to track progress during training, which is shown in Fig. \ref{fig:decomp_perform} (d) and (e). 
The qualitative comparison demonstrated that the target region was well-formed in the early epoch using HL, suggesting the effectiveness of HL. 
A summary of the experimental results for all backbones is shown in Table \ref{tab:ablation}.

\subsubsection{BMD estimation accuracy} 
\begin{figure*}[!t]
\centering
\includegraphics[width=.95\textwidth]{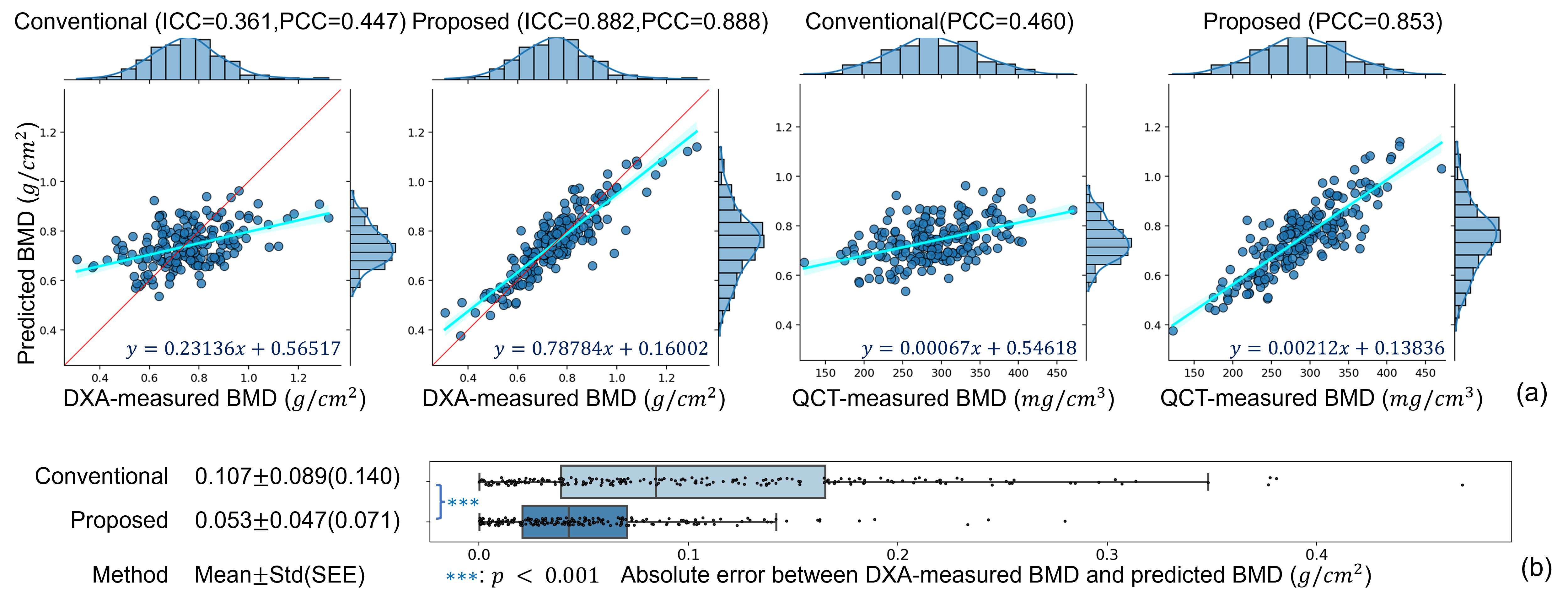}
\caption{Results of BMD estimation using a limited dataset.
(a) Correlation of the predicted BMD with DXA-BMD and QCT-BMD.
(b) Boxplot of the absolute error (AE) of the predicted BMD. The BMD predicted using the proposed method clearly shows a higher correlation with DXA-measured and QCT-measured BMDs and smaller absolute errors.}
\label{fig:200_dxabmd}
\end{figure*}

A comparison of the BMD estimation performance between the conventional method and the proposed HRFormer Generator with HL is shown in Fig. \ref{fig:200_dxabmd}
The proposed method achieved high PCC of 0.882 and 0.888 (compared to 0.361 and 0.447), respectively, demonstrating the effectiveness of the estimation strategy of the proposed method that extracts the density distribution of the target region of the bone. 
We evaluated 13 more cases for which repeated X-ray images (acquired in the standing and supine positions on the same day) were available.
The average coefficient of variation was 3.06\% $\pm$ 3.22\% when the best model, HRFormer with HL, was used.

\section{Discussion}
\subsection{Generalizability}
We showed the robustness of the proposed method against pose variation, image compression, and uncalibrated CT toward clinical application.
As shown in Fig. \ref{fig:315_pose_bmd_result}, training with fewer poses did not result in significant degradation even during evaluations with X-ray images of unseen poses.
% The dense comparison between pose prediction in Fig. \ref{fig:315_pose_to_pose_bmd} suggested the high consistency against pose variation for X-ray images of seen and unseen poses.
The dense comparison between predictions from different poses suggested the high consistency against pose variations for X-ray images of seen and unseen poses.
In the image compression experiment, our method showed high robustness against image compression.
The performance was stable unless the X-ray images were heavily compressed, as shown in Fig. \ref{fig:compress_plots}.
Training our method with uncalibrated CT achieved PCC of 0.872 and 0.922 for DXA-BMD and QCT-BMD estimation, respectively, which are slightly different from training with QCT as shown in Table \ref{tab:summary_bmd_est}.
Our results suggested that the calibration may not be needed as long as CT images are taken using the same type of scanner.

\subsection{Difference in DXA-BMD and QCT-BMD estimation}
We noticed the performance gap between estimating DXA-BMD and QCT-BMD.
When deriving BMD from GT-PF-DRR, the performance for DXA-BMD, which achieved a PCC of 0.930, is much higher than that for QCT-BMD, which achieved a PCC of 0.883.
However, our decomposition-guided regression method showed the opposite by achieving a PCC of 0.880 and 0.920 for DXA-BMD and QCT-BMD, respectively in Fig. \ref{fig:315_bmd_result}.
The 2D modalities such as DXA and PF-DRR are projected from 3D spaces, losing partial information of posterior parts.
The average intensity of PF-DRR may be better to fit DXA-BMD because PF-DRR and DXA are in the same 2D space.
However, one may need to consider the information loss that occurs during the projection when using DXA-BMD as the ground truth.
On the other hand, the QCT-BMD, which is derived via a 3D modality, appears to be easier to regress by models not only for our method but also the conventional method we compared with.
QCT-BMD could serve as additional supervision for potential improvement in DXA-BMD estimation.
Furthermore, this phenomenon may suggest the model's ability to infer 3D information from a 2D modality, encouraging us to recover the 3D spatial density of bone from a plain X-ray image in the future.

\subsection{Learning and predicting 2D BMD distribution}
A unique feature of our method is that it produces the pixel-wise BMD estimation, representing the 2D BMD distribution of the PF bone in addition to the predicted DXA-BMD in stage 2, improving explainability.
Indeed, one can tell that the predicted BMD is incorrect before knowing the predicted BMD value if the model fails to produce a reasonable PF-DRR, which inspires us to develop a failure detection mechanism for actual clinical applications. 
This ability was obtained by learning the mapping from an X-ray image to PF-DRR, suggesting better use of the texture pattern inside the bone, which is potentially helpful for the fracture risk prediction, as it has been shown to be factorized by the texture of the bone\citep{dong_random_2015,farzi_quantitating_2022}.

\subsection{Method limitations}
Although the proposed method demonstrated high performance and robustness in limited datasets, the large-error source is unclear, and the performance on a large-scale dataset was not evaluated.
We will collect more data and validate the method using larger datasets collected from different centers, involving more variation.
The error source analysis showed that hip disease in patients resulted in erroneous segmentation of the PF region, which introduced estimation error. We will improve the segmentation processing pipeline to handle the diseased hip more robustly.
Although our method can localize the proximal femur region from an X-ray image, detecting key points is unavailable. We will incorporate key points detection, which could be useful in actual applications and potentially improve the model’s understanding of the patient’s hip joint pose.
Our method used 2D-3D registration, which may limit the application.
We will investigate the performance with unpaired decomposition, which does not require registration.
Furthermore, we will generalize our method to other target structures, such as the spine.

\section{Summary}
We developed a technique for estimating BMD from a plain X-ray image through decomposition learning using CT, which additionally provides the spatial density distribution of the target bone.
Our method disentangles the soft tissue and bone from an X-ray image to improve the BMD estimation accuracy, efficiently leveraging information from a small number of datasets consisting of X-ray images, BMD values, and CT of the same patient.
We first used the proposed HL framework to train a model to decompose an X-ray image into PF-DRR, then trained a regressor to estimate BMD from the representation learned via decomposition learning.
Our method showed significant performance on DXA-BMD estimation and QCT-BMD estimation tasks with limited datasets.
The proposed HL framework stabilized the decomposition training and improved decomposition accuracy as shown in the ablation study.
Furthermore, we demonstrated the high robustness of the proposed method of BMD estimation with the validations of multi-pose X-ray images, image compression, and uncalibrated CT, demonstrating the high potential for opportunistic screening in clinical practice.

\section*{Acknowledgments}
This work was funded by MEXT/JSPS KAKENHI (19H01176, 20H04550, 21K16655).

\section*{Code availability}
The source code of our method is available from the authors (gu.yi.gu4@is.naist.jp, otake@is.naist.jp, yoshi@is.naist.jp) upon reasonable request for research activity.

\bibliographystyle{model2-names.bst}\biboptions{authoryear}
\bibliography{refs}

\begin{thebibliography}{70}
\expandafter\ifx\csname natexlab\endcsname\relax\def\natexlab#1{#1}\fi
\providecommand{\url}[1]{\texttt{#1}}
\providecommand{\href}[2]{#2}
\providecommand{\path}[1]{#1}
\providecommand{\DOIprefix}{doi:}
\providecommand{\ArXivprefix}{arXiv:}
\providecommand{\URLprefix}{URL: }
\providecommand{\Pubmedprefix}{pmid:}
\providecommand{\doi}[1]{\href{http://dx.doi.org/#1}{\path{#1}}}
\providecommand{\Pubmed}[1]{\href{pmid:#1}{\path{#1}}}
\providecommand{\bibinfo}[2]{#2}
\ifx\xfnm\relax \def\xfnm[#1]{\unskip,\space#1}\fi
%Type = Article
\bibitem[{Aggarwal et~al.(2021)Aggarwal, Maslen, Abel, Bhattacharya, Bromiley,
  Clark, Compston, Crabtree, Gregory, Kariki, Harvey, Ward and
  Poole}]{aggarwal_opportunistic_2021}
\bibinfo{author}{Aggarwal, V.}, \bibinfo{author}{Maslen, C.},
  \bibinfo{author}{Abel, R.L.}, \bibinfo{author}{Bhattacharya, P.},
  \bibinfo{author}{Bromiley, P.A.}, \bibinfo{author}{Clark, E.M.},
  \bibinfo{author}{Compston, J.E.}, \bibinfo{author}{Crabtree, N.},
  \bibinfo{author}{Gregory, J.S.}, \bibinfo{author}{Kariki, E.P.},
  \bibinfo{author}{Harvey, N.C.}, \bibinfo{author}{Ward, K.A.},
  \bibinfo{author}{Poole, K.E.S.}, \bibinfo{year}{2021}.
\newblock \bibinfo{title}{Opportunistic diagnosis of osteoporosis, fragile bone
  strength and vertebral fractures from routine {CT} scans; a review of
  approved technology systems and pathways to implementation}.
\newblock \bibinfo{journal}{Ther. Adv. Musculoskelet. Dis.}
  \bibinfo{volume}{13}, \bibinfo{pages}{1759720X211024029}.
\newblock \DOIprefix\doi{10.1177/1759720X211024029}.
%Type = Inproceedings
\bibitem[{Alshammari et~al.(2022)Alshammari, Wang, Ramanan and
  Kong}]{alshammari_long_2022}
\bibinfo{author}{Alshammari, S.}, \bibinfo{author}{Wang, Y.X.},
  \bibinfo{author}{Ramanan, D.}, \bibinfo{author}{Kong, S.},
  \bibinfo{year}{2022}.
\newblock \bibinfo{title}{Long- {Tailed} {Recognition} via {Weight}
  {Balancing}}, in: \bibinfo{booktitle}{2022 {IEEE}/{CVF} {Conference} on
  {Computer} {Vision} and {Pattern} {Recognition} ({CVPR})},
  \bibinfo{publisher}{IEEE}, \bibinfo{address}{New Orleans, LA, USA}. pp.
  \bibinfo{pages}{6887--6897}.
\newblock \URLprefix \url{https://ieeexplore.ieee.org/document/9880440/},
  \DOIprefix\doi{10.1109/CVPR52688.2022.00677}.
%Type = Article
\bibitem[{Bengio et~al.(2013)Bengio, Courville and
  Vincent}]{bengio_representation_2013}
\bibinfo{author}{Bengio, Y.}, \bibinfo{author}{Courville, A.},
  \bibinfo{author}{Vincent, P.}, \bibinfo{year}{2013}.
\newblock \bibinfo{title}{Representation {Learning}: {A} {Review} and {New}
  {Perspectives}}.
\newblock \bibinfo{journal}{IEEE Transactions on Pattern Analysis and Machine
  Intelligence} \bibinfo{volume}{35}, \bibinfo{pages}{1798--1828}.
\newblock \DOIprefix\doi{10.1109/TPAMI.2013.50}.
%Type = Article
\bibitem[{Blake and Fogelman(2007)}]{blake_role_2007}
\bibinfo{author}{Blake, G.M.}, \bibinfo{author}{Fogelman, I.},
  \bibinfo{year}{2007}.
\newblock \bibinfo{title}{Role of dual-energy {X}-ray absorptiometry in the
  diagnosis and treatment of osteoporosis}.
\newblock \bibinfo{journal}{J. Clin. Densitom.} \bibinfo{volume}{10},
  \bibinfo{pages}{102--110}.
\newblock \DOIprefix\doi{10.1016/j.jocd.2006.11.001}.
%Type = Inproceedings
\bibitem[{Chen and He(2021)}]{chen_exploring_2021}
\bibinfo{author}{Chen, X.}, \bibinfo{author}{He, K.}, \bibinfo{year}{2021}.
\newblock \bibinfo{title}{Exploring {Simple} {Siamese} {Representation}
  {Learning}}, in: \bibinfo{booktitle}{in Proc. 2021 CVPR}, pp.
  \bibinfo{pages}{15750--15758}.
%Type = Article
\bibitem[{Choi et~al.(2012)Choi, Oh, Kim, Lee and Chung}]{choi_prevalence_2012}
\bibinfo{author}{Choi, Y.J.}, \bibinfo{author}{Oh, H.J.}, \bibinfo{author}{Kim,
  D.J.}, \bibinfo{author}{Lee, Y.}, \bibinfo{author}{Chung, Y.S.},
  \bibinfo{year}{2012}.
\newblock \bibinfo{title}{The prevalence of osteoporosis in {Korean} adults
  aged 50 years or older and the higher diagnosis rates in women who were
  beneficiaries of a national screening program: {The} {Korea} {National}
  {Health} and {Nutrition} {Examination} {Survey} 2008–2009}.
\newblock \bibinfo{journal}{J. Bone Miner. Res.} \bibinfo{volume}{27},
  \bibinfo{pages}{1879--1886}.
%Type = Article
\bibitem[{Compston et~al.(2019)Compston, McClung and
  Leslie}]{compston_osteoporosis_2019}
\bibinfo{author}{Compston, J.E.}, \bibinfo{author}{McClung, M.R.},
  \bibinfo{author}{Leslie, W.D.}, \bibinfo{year}{2019}.
\newblock \bibinfo{title}{Osteoporosis}.
\newblock \bibinfo{journal}{Lancet (London, England)} \bibinfo{volume}{393},
  \bibinfo{pages}{364--376}.
%Type = Inproceedings
\bibitem[{Dai et~al.(2021)Dai, Liu, Le and Tan}]{dai_coatnet_2021}
\bibinfo{author}{Dai, Z.}, \bibinfo{author}{Liu, H.}, \bibinfo{author}{Le,
  Q.V.}, \bibinfo{author}{Tan, M.}, \bibinfo{year}{2021}.
\newblock \bibinfo{title}{{CoAtNet}: Marrying convolution and attention for all
  data sizes}, in: \bibinfo{booktitle}{NeurIPS}, pp.
  \bibinfo{pages}{3965--3977}.
%Type = Inproceedings
\bibitem[{Devlin et~al.(2019)Devlin, Chang, Lee and
  Toutanova}]{devlin_bert_2019}
\bibinfo{author}{Devlin, J.}, \bibinfo{author}{Chang, M.W.},
  \bibinfo{author}{Lee, K.}, \bibinfo{author}{Toutanova, K.},
  \bibinfo{year}{2019}.
\newblock \bibinfo{title}{{BERT}: {Pre}-training of {Deep} {Bidirectional}
  {Transformers} for {Language} {Understanding}}, in:
  \bibinfo{booktitle}{Proceedings of the 2019 {Conference} of the {North}
  {American} {Chapter} of the {Association} for {Computational} {Linguistics}:
  {Human} {Language} {Technologies}, {Volume} 1 ({Long} and {Short} {Papers})},
  pp. \bibinfo{pages}{4171--4186}.
\newblock \DOIprefix\doi{10.18653/v1/N19-1423}.
%Type = Inproceedings
\bibitem[{Donahue and Simonyan(2019)}]{donahue_large_2019}
\bibinfo{author}{Donahue, J.}, \bibinfo{author}{Simonyan, K.},
  \bibinfo{year}{2019}.
\newblock \bibinfo{title}{Large {Scale} {Adversarial} {Representation}
  {Learning}}, in: \bibinfo{booktitle}{Advances in {Neural} {Information}
  {Processing} {Systems}}.
%Type = Article
\bibitem[{Dong et~al.(2015)Dong, Pinninti, Lowe, Cussen, Ballard, Paolo and
  Shirvaikar}]{dong_random_2015}
\bibinfo{author}{Dong, X.N.}, \bibinfo{author}{Pinninti, R.},
  \bibinfo{author}{Lowe, T.}, \bibinfo{author}{Cussen, P.},
  \bibinfo{author}{Ballard, J.E.}, \bibinfo{author}{Paolo, D.D.},
  \bibinfo{author}{Shirvaikar, M.}, \bibinfo{year}{2015}.
\newblock \bibinfo{title}{Random field assessment of inhomogeneous bone mineral
  density from {DXA} scans can enhance the differentiation between
  postmenopausal women with and without hip fractures}.
\newblock \bibinfo{journal}{J Biomech} \bibinfo{volume}{48},
  \bibinfo{pages}{1043--1051}.
\newblock \URLprefix
  \url{https://www.ncbi.nlm.nih.gov/pmc/articles/PMC4380795/},
  \DOIprefix\doi{10.1016/j.jbiomech.2015.01.030}.
%Type = Article
\bibitem[{Engelke et~al.(2008)Engelke, Adams, Armbrecht, Augat, Bogado,
  Bouxsein, Felsenberg, Ito, Prevrhal, Hans and
  Lewiecki}]{engelke_clinical_2008}
\bibinfo{author}{Engelke, K.}, \bibinfo{author}{Adams, J.E.},
  \bibinfo{author}{Armbrecht, G.}, \bibinfo{author}{Augat, P.},
  \bibinfo{author}{Bogado, C.E.}, \bibinfo{author}{Bouxsein, M.L.},
  \bibinfo{author}{Felsenberg, D.}, \bibinfo{author}{Ito, M.},
  \bibinfo{author}{Prevrhal, S.}, \bibinfo{author}{Hans, D.B.},
  \bibinfo{author}{Lewiecki, E.M.}, \bibinfo{year}{2008}.
\newblock \bibinfo{title}{Clinical use of quantitative computed tomography and
  peripheral quantitative computed tomography in the management of osteoporosis
  in adults: The 2007 iscd official positions}.
\newblock \bibinfo{journal}{J. Clin. Densitom.} \bibinfo{volume}{11},
  \bibinfo{pages}{123--162}.
%Type = Article
\bibitem[{Eslami et~al.(2020)Eslami, Tabarestani, Albarqouni, Adeli, Navab and
  Adjouadi}]{eslami_image--images_2020}
\bibinfo{author}{Eslami, M.}, \bibinfo{author}{Tabarestani, S.},
  \bibinfo{author}{Albarqouni, S.}, \bibinfo{author}{Adeli, E.},
  \bibinfo{author}{Navab, N.}, \bibinfo{author}{Adjouadi, M.},
  \bibinfo{year}{2020}.
\newblock \bibinfo{title}{Image-to-images translation for multi-task organ
  segmentation and bone suppression in chest x-ray radiography}.
\newblock \bibinfo{journal}{IEEE Trans. Med. Imaging} \bibinfo{volume}{39},
  \bibinfo{pages}{2553--2565}.
%Type = Article
\bibitem[{Farzi et~al.(2022)Farzi, Pozo, McCloskey, Eastell, Harvey, Frangi and
  Wilkinson}]{farzi_quantitating_2022}
\bibinfo{author}{Farzi, M.}, \bibinfo{author}{Pozo, J.M.},
  \bibinfo{author}{McCloskey, E.}, \bibinfo{author}{Eastell, R.},
  \bibinfo{author}{Harvey, N.C.}, \bibinfo{author}{Frangi, A.F.},
  \bibinfo{author}{Wilkinson, J.M.}, \bibinfo{year}{2022}.
\newblock \bibinfo{title}{Quantitating {Age}-{Related} {BMD} {Textural}
  {Variation} from {DXA} {Region}-{Free}-{Analysis}: {A} {Study} of {Hip}
  {Fracture} {Prediction} in {Three} {Cohorts}}.
\newblock \bibinfo{journal}{Journal of Bone and Mineral Research}
  \bibinfo{volume}{37}, \bibinfo{pages}{1679--1688}.
\newblock \URLprefix
  \url{https://onlinelibrary.wiley.com/doi/abs/10.1002/jbmr.4638},
  \DOIprefix\doi{10.1002/jbmr.4638}.
%Type = Inproceedings
\bibitem[{Gu et~al.(2022)Gu, Otake, Uemura, Soufi, Takao, Sugano and
  Sato}]{gu_bmd-gan_2022}
\bibinfo{author}{Gu, Y.}, \bibinfo{author}{Otake, Y.}, \bibinfo{author}{Uemura,
  K.}, \bibinfo{author}{Soufi, M.}, \bibinfo{author}{Takao, M.},
  \bibinfo{author}{Sugano, N.}, \bibinfo{author}{Sato, Y.},
  \bibinfo{year}{2022}.
\newblock \bibinfo{title}{{BMD}-{GAN}: Bone mineral density estimation using
  x-ray image decomposition into projections of bone-segmented quantitative
  computed tomography using hierarchical learning}, in: \bibinfo{editor}{Wang,
  L.}, \bibinfo{editor}{Dou, Q.}, \bibinfo{editor}{Fletcher, P.T.},
  \bibinfo{editor}{Speidel, S.}, \bibinfo{editor}{Li, S.} (Eds.),
  \bibinfo{booktitle}{MICCAI}, \bibinfo{publisher}{Springer Nature
  Switzerland}, \bibinfo{address}{Cham}. pp. \bibinfo{pages}{644--654}.
%Type = Inproceedings
\bibitem[{Hiasa et~al.(2018)Hiasa, Otake, Takao, Matsuoka, Takashima, Carass,
  Prince, Sugano and Sato}]{hiasa_cross_2018}
\bibinfo{author}{Hiasa, Y.}, \bibinfo{author}{Otake, Y.},
  \bibinfo{author}{Takao, M.}, \bibinfo{author}{Matsuoka, T.},
  \bibinfo{author}{Takashima, K.}, \bibinfo{author}{Carass, A.},
  \bibinfo{author}{Prince, J.L.}, \bibinfo{author}{Sugano, N.},
  \bibinfo{author}{Sato, Y.}, \bibinfo{year}{2018}.
\newblock \bibinfo{title}{Cross-{Modality} {Image} {Synthesis} from {Unpaired}
  {Data} {Using} {CycleGAN}}, in: \bibinfo{editor}{Gooya, A.},
  \bibinfo{editor}{Goksel, O.}, \bibinfo{editor}{Oguz, I.},
  \bibinfo{editor}{Burgos, N.} (Eds.), \bibinfo{booktitle}{MICCAIW Simulation
  and {Synthesis} in {Medical} {Imaging}}, \bibinfo{publisher}{Springer
  International Publishing}, \bibinfo{address}{Cham}. pp.
  \bibinfo{pages}{31--41}.
\newblock \DOIprefix\doi{10.1007/978-3-030-00536-8_4}.
%Type = Article
\bibitem[{Hiasa et~al.(2020)Hiasa, Otake, Takao, Ogawa, Sugano and
  Sato}]{hiasa_automated_2020}
\bibinfo{author}{Hiasa, Y.}, \bibinfo{author}{Otake, Y.},
  \bibinfo{author}{Takao, M.}, \bibinfo{author}{Ogawa, T.},
  \bibinfo{author}{Sugano, N.}, \bibinfo{author}{Sato, Y.},
  \bibinfo{year}{2020}.
\newblock \bibinfo{title}{Automated muscle segmentation from clinical {CT}
  using {Bayesian} {U}-{Net} for personalized musculoskeletal modeling}.
\newblock \bibinfo{journal}{IEEE Trans. Med. Imaging} \bibinfo{volume}{39},
  \bibinfo{pages}{1030--1040}.
%Type = Article
\bibitem[{Ho et~al.(2021)Ho, Chen, Fan, Kuo, Yen, Liu and
  Pei}]{ho_application_2021}
\bibinfo{author}{Ho, C.S.}, \bibinfo{author}{Chen, Y.P.}, \bibinfo{author}{Fan,
  T.Y.}, \bibinfo{author}{Kuo, C.F.}, \bibinfo{author}{Yen, T.Y.},
  \bibinfo{author}{Liu, Y.C.}, \bibinfo{author}{Pei, Y.C.},
  \bibinfo{year}{2021}.
\newblock \bibinfo{title}{Application of deep learning neural network in
  predicting bone mineral density from plain {X}-ray radiography}.
\newblock \bibinfo{journal}{Arch. Osteoporos.} \bibinfo{volume}{16},
  \bibinfo{pages}{153}.
%Type = Inproceedings
\bibitem[{Hoyer et~al.(2022)Hoyer, Dai and Van~Gool}]{hoyer_daformer_2022}
\bibinfo{author}{Hoyer, L.}, \bibinfo{author}{Dai, D.},
  \bibinfo{author}{Van~Gool, L.}, \bibinfo{year}{2022}.
\newblock \bibinfo{title}{{DAFormer}: Improving network architectures and
  training strategies for domain-adaptive semantic segmentation}, in:
  \bibinfo{booktitle}{CVPR}, pp. \bibinfo{pages}{9924--9935}.
%Type = Article
\bibitem[{Hsieh et~al.(2021)Hsieh, Zheng, Lin, Mei, Lu, Li, Chen, Wang, Zhou,
  Wang, Xie, Xiao, Miao and Kuo}]{hsieh_automated_2021}
\bibinfo{author}{Hsieh, C.I.}, \bibinfo{author}{Zheng, K.},
  \bibinfo{author}{Lin, C.}, \bibinfo{author}{Mei, L.}, \bibinfo{author}{Lu,
  L.}, \bibinfo{author}{Li, W.}, \bibinfo{author}{Chen, F.P.},
  \bibinfo{author}{Wang, Y.}, \bibinfo{author}{Zhou, X.},
  \bibinfo{author}{Wang, F.}, \bibinfo{author}{Xie, G.}, \bibinfo{author}{Xiao,
  J.}, \bibinfo{author}{Miao, S.}, \bibinfo{author}{Kuo, C.F.},
  \bibinfo{year}{2021}.
\newblock \bibinfo{title}{Automated bone mineral density prediction and
  fracture risk assessment using plain radiographs via deep learning}.
\newblock \bibinfo{journal}{Nat. Communi.} \bibinfo{volume}{12},
  \bibinfo{pages}{5472}.
%Type = Article
\bibitem[{Iki et~al.(2001)Iki, Kagamimori, Kagawa, Matsuzaki, Yoneshima, Marumo
  and {for JPOS Study Group}}]{iki_bone_2001}
\bibinfo{author}{Iki, M.}, \bibinfo{author}{Kagamimori, S.},
  \bibinfo{author}{Kagawa, Y.}, \bibinfo{author}{Matsuzaki, T.},
  \bibinfo{author}{Yoneshima, H.}, \bibinfo{author}{Marumo, F.},
  \bibinfo{author}{{for JPOS Study Group}}, \bibinfo{year}{2001}.
\newblock \bibinfo{title}{Bone mineral density of the spine, hip and distal
  forearm in representative samples of the japanese female population:
  {Japanese} population-based osteoporosis ({JPOS}) study}.
\newblock \bibinfo{journal}{Osteoporos. Int.} \bibinfo{volume}{12},
  \bibinfo{pages}{529--537}.
%Type = Inproceedings
\bibitem[{Isola et~al.(2017)Isola, Zhu, Zhou and
  Efros}]{isola_image--image_2017}
\bibinfo{author}{Isola, P.}, \bibinfo{author}{Zhu, J.Y.},
  \bibinfo{author}{Zhou, T.}, \bibinfo{author}{Efros, A.A.},
  \bibinfo{year}{2017}.
\newblock \bibinfo{title}{Image-to-image translation with conditional
  adversarial networks}, in: \bibinfo{booktitle}{in Proc. 2017 CVPR}, pp.
  \bibinfo{pages}{5967--5976}.
%Type = Article
\bibitem[{Jang et~al.(2021)Jang, Choi, Kim, Chang, Yoon and
  Kim}]{jang_prediction_2021}
\bibinfo{author}{Jang, R.}, \bibinfo{author}{Choi, J.H.}, \bibinfo{author}{Kim,
  N.}, \bibinfo{author}{Chang, J.S.}, \bibinfo{author}{Yoon, P.W.},
  \bibinfo{author}{Kim, C.H.}, \bibinfo{year}{2021}.
\newblock \bibinfo{title}{Prediction of osteoporosis from simple hip
  radiography using deep learning algorithm}.
\newblock \bibinfo{journal}{Sci. Rep.} \bibinfo{volume}{11},
  \bibinfo{pages}{19997}.
\newblock \DOIprefix\doi{10.1038/s41598-021-99549-6}.
%Type = Article
\bibitem[{Kohn et~al.(2016)Kohn, Sassoon and
  Fernando}]{kohn_classifications_2016}
\bibinfo{author}{Kohn, M.D.}, \bibinfo{author}{Sassoon, A.A.},
  \bibinfo{author}{Fernando, N.D.}, \bibinfo{year}{2016}.
\newblock \bibinfo{title}{Classifications in {Brief}: {Kellgren}-{Lawrence}
  {Classification} of {Osteoarthritis}}.
\newblock \bibinfo{journal}{Clin Orthop Relat Res} \bibinfo{volume}{474},
  \bibinfo{pages}{1886--1893}.
\newblock \DOIprefix\doi{10.1007/s11999-016-4732-4}.
%Type = Article
\bibitem[{Kröger et~al.(1992)Kröger, Kotaniemi, Vainio and
  Alhava}]{kroger_bone_1992}
\bibinfo{author}{Kröger, H.}, \bibinfo{author}{Kotaniemi, A.},
  \bibinfo{author}{Vainio, P.}, \bibinfo{author}{Alhava, E.},
  \bibinfo{year}{1992}.
\newblock \bibinfo{title}{Bone densitometry of the spine and femur in children
  by dual-energy x-ray absorptiometry}.
\newblock \bibinfo{journal}{Bone Miner.} \bibinfo{volume}{17},
  \bibinfo{pages}{75--85}.
%Type = Article
\bibitem[{Kung et~al.(2013)Kung, Fan, Xu, Xia, Park, Kim, Chan, Lee, Koh,
  Soong, Soontrapa, Songpatanasilp, Turajane, Yates and
  Sen}]{kung_factors_2013}
\bibinfo{author}{Kung, A.W.}, \bibinfo{author}{Fan, T.}, \bibinfo{author}{Xu,
  L.}, \bibinfo{author}{Xia, W.B.}, \bibinfo{author}{Park, I.H.},
  \bibinfo{author}{Kim, H.S.}, \bibinfo{author}{Chan, S.P.},
  \bibinfo{author}{Lee, J.K.}, \bibinfo{author}{Koh, L.},
  \bibinfo{author}{Soong, Y.K.}, \bibinfo{author}{Soontrapa, S.},
  \bibinfo{author}{Songpatanasilp, T.}, \bibinfo{author}{Turajane, T.},
  \bibinfo{author}{Yates, M.}, \bibinfo{author}{Sen, S.}, \bibinfo{year}{2013}.
\newblock \bibinfo{title}{Factors influencing diagnosis and treatment of
  osteoporosis after a fragility fracture among postmenopausal women in {Asian}
  countries: a retrospective study}.
\newblock \bibinfo{journal}{BMC Women's Health} \bibinfo{volume}{13},
  \bibinfo{pages}{7}.
%Type = Article
\bibitem[{Lewiecki et~al.(2016)Lewiecki, Binkley, Morgan, Shuhart, Camargos,
  Carey, Gordon, Jankowski, Lee, Leslie and {International Society for Clinical
  Densitometry}}]{lewiecki_best_2016}
\bibinfo{author}{Lewiecki, E.M.}, \bibinfo{author}{Binkley, N.},
  \bibinfo{author}{Morgan, S.L.}, \bibinfo{author}{Shuhart, C.R.},
  \bibinfo{author}{Camargos, B.M.}, \bibinfo{author}{Carey, J.J.},
  \bibinfo{author}{Gordon, C.M.}, \bibinfo{author}{Jankowski, L.G.},
  \bibinfo{author}{Lee, J.K.}, \bibinfo{author}{Leslie, W.D.},
  \bibinfo{author}{{International Society for Clinical Densitometry}},
  \bibinfo{year}{2016}.
\newblock \bibinfo{title}{Best {Practices} for {Dual}-{Energy} {X}-ray
  {Absorptiometry} {Measurement} and {Reporting}: {International} {Society} for
  {Clinical} {Densitometry} {Guidance}}.
\newblock \bibinfo{journal}{J Clin Densitom} \bibinfo{volume}{19},
  \bibinfo{pages}{127--140}.
\newblock \DOIprefix\doi{10.1016/j.jocd.2016.03.003}.
%Type = Article
\bibitem[{Li et~al.(2021)Li, Xiao, Quarles and Li}]{li_osteoporosis_2021}
\bibinfo{author}{Li, H.}, \bibinfo{author}{Xiao, Z.}, \bibinfo{author}{Quarles,
  L.D.}, \bibinfo{author}{Li, W.}, \bibinfo{year}{2021}.
\newblock \bibinfo{title}{Osteoporosis: Mechanism, molecular target and current
  status on drug development}.
\newblock \bibinfo{journal}{Curr. Med. Chem.} \bibinfo{volume}{28},
  \bibinfo{pages}{1489--1507}.
%Type = Inproceedings
\bibitem[{Li et~al.(2022)Li, Cao, Yuan, Fan, Yang, Feris, Indyk and
  Katabi}]{li_targeted_2022}
\bibinfo{author}{Li, T.}, \bibinfo{author}{Cao, P.}, \bibinfo{author}{Yuan,
  Y.}, \bibinfo{author}{Fan, L.}, \bibinfo{author}{Yang, Y.},
  \bibinfo{author}{Feris, R.}, \bibinfo{author}{Indyk, P.},
  \bibinfo{author}{Katabi, D.}, \bibinfo{year}{2022}.
\newblock \bibinfo{title}{Targeted {Supervised} {Contrastive} {Learning} for
  {Long}-{Tailed} {Recognition}}, in: \bibinfo{booktitle}{2022 {IEEE}/{CVF}
  {Conference} on {Computer} {Vision} and {Pattern} {Recognition} ({CVPR})},
  \bibinfo{publisher}{IEEE}, \bibinfo{address}{New Orleans, LA, USA}. pp.
  \bibinfo{pages}{6908--6918}.
\newblock \URLprefix \url{https://ieeexplore.ieee.org/document/9878544/},
  \DOIprefix\doi{10.1109/CVPR52688.2022.00679}.
%Type = Article
\bibitem[{Liu et~al.(2021)Liu, Fan, Zhang, Zhou, Xiao, Geng and
  Shen}]{liu_incomplete_2021}
\bibinfo{author}{Liu, Y.}, \bibinfo{author}{Fan, L.}, \bibinfo{author}{Zhang,
  C.}, \bibinfo{author}{Zhou, T.}, \bibinfo{author}{Xiao, Z.},
  \bibinfo{author}{Geng, L.}, \bibinfo{author}{Shen, D.}, \bibinfo{year}{2021}.
\newblock \bibinfo{title}{Incomplete multi-modal representation learning for
  {Alzheimer}’s disease diagnosis}.
\newblock \bibinfo{journal}{Medical Image Analysis} \bibinfo{volume}{69},
  \bibinfo{pages}{101953}.
\newblock \DOIprefix\doi{10.1016/j.media.2020.101953}.
%Type = Article
\bibitem[{Liu et~al.(2019)Liu, Zhang, Cai, Chen, Yun, Feng and
  Yang}]{liu_automatic_2019}
\bibinfo{author}{Liu, Y.}, \bibinfo{author}{Zhang, X.}, \bibinfo{author}{Cai,
  G.}, \bibinfo{author}{Chen, Y.}, \bibinfo{author}{Yun, Z.},
  \bibinfo{author}{Feng, Q.}, \bibinfo{author}{Yang, W.}, \bibinfo{year}{2019}.
\newblock \bibinfo{title}{Automatic delineation of ribs and clavicles in chest
  radiographs using fully convolutional {DenseNets}}.
\newblock \bibinfo{journal}{Comput. Meth. Prog. Bio.} \bibinfo{volume}{180},
  \bibinfo{pages}{105014}.
%Type = Article
\bibitem[{Lorentzon(2019)}]{lorentzon_treating_2019}
\bibinfo{author}{Lorentzon, M.}, \bibinfo{year}{2019}.
\newblock \bibinfo{title}{Treating osteoporosis to prevent fractures: current
  concepts and future developments}.
\newblock \bibinfo{journal}{J. Intern. Med.} \bibinfo{volume}{285},
  \bibinfo{pages}{381--394}.
%Type = Inproceedings
\bibitem[{Loshchilov and Hutter(2017)}]{loshchilov_sgdr_2017}
\bibinfo{author}{Loshchilov, I.}, \bibinfo{author}{Hutter, F.},
  \bibinfo{year}{2017}.
\newblock \bibinfo{title}{{SGDR}: Stochastic gradient descent with warm
  restarts}, in: \bibinfo{booktitle}{ICLR}.
%Type = Inproceedings
\bibitem[{Loshchilov and Hutter(2019)}]{loshchilov_decoupled_2019}
\bibinfo{author}{Loshchilov, I.}, \bibinfo{author}{Hutter, F.},
  \bibinfo{year}{2019}.
\newblock \bibinfo{title}{Decoupled weight decay regularization}, in:
  \bibinfo{booktitle}{ICLR}.
%Type = Article
\bibitem[{Löffler et~al.(2020)Löffler, Sollmann, Mei, Valentinitsch, Noël,
  Kirschke and Baum}]{loffler_X-ray-based_2020}
\bibinfo{author}{Löffler, M.}, \bibinfo{author}{Sollmann, N.},
  \bibinfo{author}{Mei, K.}, \bibinfo{author}{Valentinitsch, A.},
  \bibinfo{author}{Noël, P.}, \bibinfo{author}{Kirschke, J.},
  \bibinfo{author}{Baum, T.}, \bibinfo{year}{2020}.
\newblock \bibinfo{title}{X-ray-based quantitative osteoporosis imaging at the
  spine}.
\newblock \bibinfo{journal}{Osteoporos. Int.} \bibinfo{volume}{31},
  \bibinfo{pages}{233--250}.
%Type = Article
\bibitem[{Mazess et~al.(1990)Mazess, Barden, Bisek and
  Hanson}]{mazess_dual-energy_1990}
\bibinfo{author}{Mazess, R.B.}, \bibinfo{author}{Barden, H.S.},
  \bibinfo{author}{Bisek, J.P.}, \bibinfo{author}{Hanson, J.},
  \bibinfo{year}{1990}.
\newblock \bibinfo{title}{Dual-energy x-ray absorptiometry for total-body and
  regional bone-mineral and soft-tissue composition}.
\newblock \bibinfo{journal}{Am. J. Clin. Nutr.} \bibinfo{volume}{51},
  \bibinfo{pages}{1106--1112}.
%Type = Article
\bibitem[{McCloskey et~al.(2021)McCloskey, Rathi, Heijmans, Blagden, Cortet,
  Czerwinski, Hadji, Payer, Palmer, Stad, O’Kelly and
  Papapoulos}]{mccloskey_osteoporosis_2021}
\bibinfo{author}{McCloskey, E.}, \bibinfo{author}{Rathi, J.},
  \bibinfo{author}{Heijmans, S.}, \bibinfo{author}{Blagden, M.},
  \bibinfo{author}{Cortet, B.}, \bibinfo{author}{Czerwinski, E.},
  \bibinfo{author}{Hadji, P.}, \bibinfo{author}{Payer, J.},
  \bibinfo{author}{Palmer, K.}, \bibinfo{author}{Stad, R.},
  \bibinfo{author}{O’Kelly, J.}, \bibinfo{author}{Papapoulos, S.},
  \bibinfo{year}{2021}.
\newblock \bibinfo{title}{The osteoporosis treatment gap in patients at risk of
  fracture in {European} primary care: a multi-country cross-sectional
  observational study}.
\newblock \bibinfo{journal}{Osteoporos. Int.} \bibinfo{volume}{32},
  \bibinfo{pages}{251--259}.
%Type = Article
\bibitem[{Mueller et~al.(2011)Mueller, Kutscherenko, Bartel, Vlassenbroek,
  Ourednicek and Erckenbrecht}]{mueller_phantom-less_2011}
\bibinfo{author}{Mueller, D.K.}, \bibinfo{author}{Kutscherenko, A.},
  \bibinfo{author}{Bartel, H.}, \bibinfo{author}{Vlassenbroek, A.},
  \bibinfo{author}{Ourednicek, P.}, \bibinfo{author}{Erckenbrecht, J.},
  \bibinfo{year}{2011}.
\newblock \bibinfo{title}{Phantom-less {QCT} {BMD} system as screening tool for
  osteoporosis without additional radiation}.
\newblock \bibinfo{journal}{Eur. J. Radiol.} \bibinfo{volume}{79},
  \bibinfo{pages}{375--381}.
\newblock \DOIprefix\doi{10.1016/j.ejrad.2010.02.008}.
%Type = Article
\bibitem[{Noh et~al.(2020)Noh, Yang and Jung}]{noh_molecular_2020}
\bibinfo{author}{Noh, J.Y.}, \bibinfo{author}{Yang, Y.}, \bibinfo{author}{Jung,
  H.}, \bibinfo{year}{2020}.
\newblock \bibinfo{title}{Molecular mechanisms and emerging therapeutics for
  osteoporosis}.
\newblock \bibinfo{journal}{Int. J. Mol. Sci.} \bibinfo{volume}{21},
  \bibinfo{pages}{7623}.
%Type = Inproceedings
\bibitem[{Noroozi et~al.(2017)Noroozi, Pirsiavash and
  Favaro}]{noroozi_representation_2017}
\bibinfo{author}{Noroozi, M.}, \bibinfo{author}{Pirsiavash, H.},
  \bibinfo{author}{Favaro, P.}, \bibinfo{year}{2017}.
\newblock \bibinfo{title}{Representation {Learning} by {Learning} to {Count}},
  in: \bibinfo{booktitle}{in Proc. 2017 ICCV}, pp. \bibinfo{pages}{5898--5906}.
%Type = Article
\bibitem[{O'Malley et~al.(2011)O'Malley, Johnston, Lenhart, Cherkowski, Palmer
  and Morgan}]{omalley_trends_2011}
\bibinfo{author}{O'Malley, C.D.}, \bibinfo{author}{Johnston, S.S.},
  \bibinfo{author}{Lenhart, G.}, \bibinfo{author}{Cherkowski, G.},
  \bibinfo{author}{Palmer, L.}, \bibinfo{author}{Morgan, S.L.},
  \bibinfo{year}{2011}.
\newblock \bibinfo{title}{Trends in dual-energy x-ray absorptiometry in the
  {United} {States}, 2000–2009}.
\newblock \bibinfo{journal}{J. Clin. Densitom.} \bibinfo{volume}{14},
  \bibinfo{pages}{100--107}.
%Type = Article
\bibitem[{Otake et~al.(2012)Otake, Armand, Armiger, Kutzer, Basafa, Kazanzides
  and Taylor}]{otake_intraoperative_2012}
\bibinfo{author}{Otake, Y.}, \bibinfo{author}{Armand, M.},
  \bibinfo{author}{Armiger, R.S.}, \bibinfo{author}{Kutzer, M.D.},
  \bibinfo{author}{Basafa, E.}, \bibinfo{author}{Kazanzides, P.},
  \bibinfo{author}{Taylor, R.H.}, \bibinfo{year}{2012}.
\newblock \bibinfo{title}{Intraoperative image-based multiview {2D}/{3D}
  registration for image-guided orthopaedic surgery: incorporation of
  fiducial-based {C}-arm tracking and {GPU}-acceleration}.
\newblock \bibinfo{journal}{IEEE Trans. Med. Imaging} \bibinfo{volume}{31},
  \bibinfo{pages}{948--962}.
%Type = Article
\bibitem[{Papaioannou et~al.(2008)Papaioannou, Kennedy, Ioannidis, Gao, Sawka,
  Goltzman, Tenenhouse, Pickard, Olszynski, Davison, Kaiser, Josse, Kreiger,
  Hanley, Prior, Brown, Anastassiades, Adachi and {CaMos Research
  Group}}]{papaioannou_osteoporosis_2008}
\bibinfo{author}{Papaioannou, A.}, \bibinfo{author}{Kennedy, C.C.},
  \bibinfo{author}{Ioannidis, G.}, \bibinfo{author}{Gao, Y.},
  \bibinfo{author}{Sawka, A.M.}, \bibinfo{author}{Goltzman, D.},
  \bibinfo{author}{Tenenhouse, A.}, \bibinfo{author}{Pickard, L.},
  \bibinfo{author}{Olszynski, W.P.}, \bibinfo{author}{Davison, K.S.},
  \bibinfo{author}{Kaiser, S.}, \bibinfo{author}{Josse, R.G.},
  \bibinfo{author}{Kreiger, N.}, \bibinfo{author}{Hanley, D.A.},
  \bibinfo{author}{Prior, J.C.}, \bibinfo{author}{Brown, J.P.},
  \bibinfo{author}{Anastassiades, T.}, \bibinfo{author}{Adachi, J.D.},
  \bibinfo{author}{{CaMos Research Group}}, \bibinfo{year}{2008}.
\newblock \bibinfo{title}{The osteoporosis care gap in men with fragility
  fractures: The {Canadian} multicentre osteoporosis study}.
\newblock \bibinfo{journal}{Osteoporos. Int.} \bibinfo{volume}{19},
  \bibinfo{pages}{581--587}.
\newblock \URLprefix \url{https://doi.org/10.1007/s00198-007-0483-0},
  \DOIprefix\doi{10.1007/s00198-007-0483-0}.
%Type = Article
\bibitem[{Penney et~al.(1998)Penney, Weese, Little, Desmedt, Hill and
  hawkes}]{penney_comparison_1998}
\bibinfo{author}{Penney, G.}, \bibinfo{author}{Weese, J.},
  \bibinfo{author}{Little, J.}, \bibinfo{author}{Desmedt, P.},
  \bibinfo{author}{Hill, D.}, \bibinfo{author}{hawkes, D.},
  \bibinfo{year}{1998}.
\newblock \bibinfo{title}{A comparison of similarity measures for use in
  2-{D}-3-{D} medical image registration}.
\newblock \bibinfo{journal}{IEEE Trans. Med. Imaging} \bibinfo{volume}{17},
  \bibinfo{pages}{586--595}.
%Type = Article
\bibitem[{Pisani et~al.(2013)Pisani, Renna, Conversano, Casciaro, Muratore,
  Quarta, Paola and Casciaro}]{pisani_screening_2013}
\bibinfo{author}{Pisani, P.}, \bibinfo{author}{Renna, M.D.},
  \bibinfo{author}{Conversano, F.}, \bibinfo{author}{Casciaro, E.},
  \bibinfo{author}{Muratore, M.}, \bibinfo{author}{Quarta, E.},
  \bibinfo{author}{Paola, M.D.}, \bibinfo{author}{Casciaro, S.},
  \bibinfo{year}{2013}.
\newblock \bibinfo{title}{Screening and early diagnosis of osteoporosis through
  {X}-ray and ultrasound based techniques}.
\newblock \bibinfo{journal}{World J. Radiol.} \bibinfo{volume}{5},
  \bibinfo{pages}{398--410}.
%Type = Article
\bibitem[{Snodgrass et~al.(2022)Snodgrass, Zou, Gruntmanis and
  Gitajn}]{snodgrass_osteoporosis_2022}
\bibinfo{author}{Snodgrass, P.}, \bibinfo{author}{Zou, A.},
  \bibinfo{author}{Gruntmanis, U.}, \bibinfo{author}{Gitajn, I.L.},
  \bibinfo{year}{2022}.
\newblock \bibinfo{title}{Osteoporosis diagnosis, management, and referral
  practice after fragility fractures}.
\newblock \bibinfo{journal}{Curr. Osteoporos. Rep.} \bibinfo{volume}{20},
  \bibinfo{pages}{163--169}.
%Type = Inproceedings
\bibitem[{Srinivas and Fleuret(2019)}]{srinivas_full-gradient_2019}
\bibinfo{author}{Srinivas, S.}, \bibinfo{author}{Fleuret, F.},
  \bibinfo{year}{2019}.
\newblock \bibinfo{title}{Full-gradient representation for neural network
  visualization}, in: \bibinfo{booktitle}{NeurIPS}.
%Type = Article
\bibitem[{Sugano et~al.(1998)Sugano, Noble, Kamaric, Salama, Ochi and
  Tullos}]{sugano_morphology_1998}
\bibinfo{author}{Sugano, N.}, \bibinfo{author}{Noble, P.C.},
  \bibinfo{author}{Kamaric, E.}, \bibinfo{author}{Salama, J.K.},
  \bibinfo{author}{Ochi, T.}, \bibinfo{author}{Tullos, H.S.},
  \bibinfo{year}{1998}.
\newblock \bibinfo{title}{The morphology of the femur in developmental
  dysplasia of the hip}.
\newblock \bibinfo{journal}{J Bone Joint Surg Br} \bibinfo{volume}{80},
  \bibinfo{pages}{711--719}.
\newblock \DOIprefix\doi{10.1302/0301-620x.80b4.8319}.
%Type = Article
\bibitem[{Suzuki et~al.(2006)Suzuki, Abe, MacMahon and
  Doi}]{suzuki_image-processing_2006}
\bibinfo{author}{Suzuki, K.}, \bibinfo{author}{Abe, H.},
  \bibinfo{author}{MacMahon, H.}, \bibinfo{author}{Doi, K.},
  \bibinfo{year}{2006}.
\newblock \bibinfo{title}{Image-processing technique for suppressing ribs in
  chest radiographs by means of massive training artificial neural network
  ({MTANN})}.
\newblock \bibinfo{journal}{IEEE Trans. Med. Imaging} \bibinfo{volume}{25},
  \bibinfo{pages}{406--416}.
%Type = Inproceedings
\bibitem[{Tseng et~al.(2021)Tseng, Jiang, Liu, Yang and
  Yang}]{tseng_regularizing_2021}
\bibinfo{author}{Tseng, H.Y.}, \bibinfo{author}{Jiang, L.},
  \bibinfo{author}{Liu, C.}, \bibinfo{author}{Yang, M.H.},
  \bibinfo{author}{Yang, W.}, \bibinfo{year}{2021}.
\newblock \bibinfo{title}{Regularizing generative adversarial networks under
  limited data}, in: \bibinfo{booktitle}{CVPR}, pp.
  \bibinfo{pages}{7917--7927}.
%Type = Article
\bibitem[{Uemura et~al.(2022)Uemura, Otake, Takao, Makino, Soufi, Iwasa, Sugano
  and Sato}]{uemura_development_2022}
\bibinfo{author}{Uemura, K.}, \bibinfo{author}{Otake, Y.},
  \bibinfo{author}{Takao, M.}, \bibinfo{author}{Makino, H.},
  \bibinfo{author}{Soufi, M.}, \bibinfo{author}{Iwasa, M.},
  \bibinfo{author}{Sugano, N.}, \bibinfo{author}{Sato, Y.},
  \bibinfo{year}{2022}.
\newblock \bibinfo{title}{Development of an open-source measurement system to
  assess the areal bone mineral density of the proximal femur from clinical
  {CT} images}.
\newblock \bibinfo{journal}{Arch. Osteoporos.} \bibinfo{volume}{17},
  \bibinfo{pages}{17}.
%Type = Article
\bibitem[{Uemura et~al.(2021)Uemura, Otake, Takao, Soufi, Kawasaki, Sugano and
  Sato}]{uemura_automated_2021}
\bibinfo{author}{Uemura, K.}, \bibinfo{author}{Otake, Y.},
  \bibinfo{author}{Takao, M.}, \bibinfo{author}{Soufi, M.},
  \bibinfo{author}{Kawasaki, A.}, \bibinfo{author}{Sugano, N.},
  \bibinfo{author}{Sato, Y.}, \bibinfo{year}{2021}.
\newblock \bibinfo{title}{Automated segmentation of an intensity calibration
  phantom in clinical {CT} images using a convolutional neural network}.
\newblock \bibinfo{journal}{Int. J. Comput. Assist. Radiol. Surg.}
  \bibinfo{volume}{16}, \bibinfo{pages}{1855--1864}.
%Type = Article
\bibitem[{Uemura et~al.(2023)Uemura, Takao, Otake, Takashima, Hamada, Ando,
  Sato and Sugano}]{uemura_effect_2023}
\bibinfo{author}{Uemura, K.}, \bibinfo{author}{Takao, M.},
  \bibinfo{author}{Otake, Y.}, \bibinfo{author}{Takashima, K.},
  \bibinfo{author}{Hamada, H.}, \bibinfo{author}{Ando, W.},
  \bibinfo{author}{Sato, Y.}, \bibinfo{author}{Sugano, N.},
  \bibinfo{year}{2023}.
\newblock \bibinfo{title}{The effect of patient positioning on measurements of
  bone mineral density of the proximal femur: a simulation study using computed
  tomographic images}.
\newblock \bibinfo{journal}{Arch Osteoporos} \bibinfo{volume}{18},
  \bibinfo{pages}{35}.
\newblock \DOIprefix\doi{10.1007/s11657-023-01225-x}.
%Type = Article
\bibitem[{Wang et~al.(2023)Wang, Zheng, Lu, Xiao, Wu, Kuo and
  Miao}]{wang_lumbar_2023}
\bibinfo{author}{Wang, F.}, \bibinfo{author}{Zheng, K.}, \bibinfo{author}{Lu,
  L.}, \bibinfo{author}{Xiao, J.}, \bibinfo{author}{Wu, M.},
  \bibinfo{author}{Kuo, C.F.}, \bibinfo{author}{Miao, S.},
  \bibinfo{year}{2023}.
\newblock \bibinfo{title}{Lumbar {Bone} {Mineral} {Density} {Estimation} {From}
  {Chest} {X}-{Ray} {Images}: {Anatomy}-{Aware} {Attentive} {Multi}-{ROI}
  {Modeling}}.
\newblock \bibinfo{journal}{IEEE Transactions on Medical Imaging}
  \bibinfo{volume}{42}, \bibinfo{pages}{257--267}.
\newblock \DOIprefix\doi{10.1109/TMI.2022.3209648}.
%Type = Article
\bibitem[{Wang et~al.(2021)Wang, Sun, Cheng, Jiang, Deng, Zhao, Liu, Mu, Tan,
  Wang, Liu and Xiao}]{wang_deep_2021}
\bibinfo{author}{Wang, J.}, \bibinfo{author}{Sun, K.}, \bibinfo{author}{Cheng,
  T.}, \bibinfo{author}{Jiang, B.}, \bibinfo{author}{Deng, C.},
  \bibinfo{author}{Zhao, Y.}, \bibinfo{author}{Liu, D.}, \bibinfo{author}{Mu,
  Y.}, \bibinfo{author}{Tan, M.}, \bibinfo{author}{Wang, X.},
  \bibinfo{author}{Liu, W.}, \bibinfo{author}{Xiao, B.}, \bibinfo{year}{2021}.
\newblock \bibinfo{title}{Deep high-resolution representation learning for
  visual recognition}.
\newblock \bibinfo{journal}{IEEE Trans. Pattern Anal. Mach. Intell.}
  \bibinfo{volume}{43}, \bibinfo{pages}{3349--3364}.
%Type = Inproceedings
\bibitem[{Wang et~al.(2018)Wang, Liu, Zhu, Tao, Kautz and
  Catanzaro}]{wang_high-resolution_2018}
\bibinfo{author}{Wang, T.C.}, \bibinfo{author}{Liu, M.Y.},
  \bibinfo{author}{Zhu, J.Y.}, \bibinfo{author}{Tao, A.},
  \bibinfo{author}{Kautz, J.}, \bibinfo{author}{Catanzaro, B.},
  \bibinfo{year}{2018}.
\newblock \bibinfo{title}{High-resolution image synthesis and semantic
  manipulation with conditional {GANs}}, in: \bibinfo{booktitle}{CVPR}, pp.
  \bibinfo{pages}{8798--8807}.
%Type = Article
\bibitem[{Ward et~al.(2020)Ward, Weber, Munns, Högler and
  Zemel}]{ward_contemporary_2020}
\bibinfo{author}{Ward, L.M.}, \bibinfo{author}{Weber, D.R.},
  \bibinfo{author}{Munns, C.F.}, \bibinfo{author}{Högler, W.},
  \bibinfo{author}{Zemel, B.S.}, \bibinfo{year}{2020}.
\newblock \bibinfo{title}{A contemporary view of the definition and diagnosis
  of osteoporosis in children and adolescents}.
\newblock \bibinfo{journal}{J. Clin. Endocrinol. Metab.} \bibinfo{volume}{105},
  \bibinfo{pages}{e2088--e2097}.
%Type = Article
\bibitem[{Whitmarsh et~al.(2011)Whitmarsh, Humbert, De~Craene, Del Rio~Barquero
  and Frangi}]{whitmarsh_reconstructing_2011}
\bibinfo{author}{Whitmarsh, T.}, \bibinfo{author}{Humbert, L.},
  \bibinfo{author}{De~Craene, M.}, \bibinfo{author}{Del Rio~Barquero, L.M.},
  \bibinfo{author}{Frangi, A.F.}, \bibinfo{year}{2011}.
\newblock \bibinfo{title}{Reconstructing the {3D} shape and bone mineral
  density distribution of the proximal femur from dual-energy {X}-ray
  absorptiometry}.
\newblock \bibinfo{journal}{IEEE Tran. Med. Imaging} \bibinfo{volume}{30},
  \bibinfo{pages}{2101--2114}.
\newblock \DOIprefix\doi{10.1109/TMI.2011.2163074}.
%Type = Inproceedings
\bibitem[{Wu et~al.(2021)Wu, Shuai, Tam and Chiu}]{wu_gradient_2021}
\bibinfo{author}{Wu, Y.L.}, \bibinfo{author}{Shuai, H.H.},
  \bibinfo{author}{Tam, Z.R.}, \bibinfo{author}{Chiu, H.Y.},
  \bibinfo{year}{2021}.
\newblock \bibinfo{title}{Gradient normalization for generative adversarial
  networks}, in: \bibinfo{booktitle}{ICCV}, pp. \bibinfo{pages}{6353--6362}.
%Type = Article
\bibitem[{Yamamoto et~al.(2020)Yamamoto, Sukegawa, Kitamura, Goto, Noda,
  Nakano, Takabatake, Kawai, Nagatsuka, Kawasaki, Furuki and
  Ozaki}]{yamamoto_deep_2020}
\bibinfo{author}{Yamamoto, N.}, \bibinfo{author}{Sukegawa, S.},
  \bibinfo{author}{Kitamura, A.}, \bibinfo{author}{Goto, R.},
  \bibinfo{author}{Noda, T.}, \bibinfo{author}{Nakano, K.},
  \bibinfo{author}{Takabatake, K.}, \bibinfo{author}{Kawai, H.},
  \bibinfo{author}{Nagatsuka, H.}, \bibinfo{author}{Kawasaki, K.},
  \bibinfo{author}{Furuki, Y.}, \bibinfo{author}{Ozaki, T.},
  \bibinfo{year}{2020}.
\newblock \bibinfo{title}{Deep learning for osteoporosis classification using
  hip radiographs and patient clinical covariates}.
\newblock \bibinfo{journal}{Biomolecules} \bibinfo{volume}{10},
  \bibinfo{pages}{E1534}.
%Type = Article
\bibitem[{Yang et~al.(2020)Yang, Shen, Liu, Dong, Zhang, Deng, Zhao and
  Deng}]{yang_road_2020}
\bibinfo{author}{Yang, T.L.}, \bibinfo{author}{Shen, H.}, \bibinfo{author}{Liu,
  A.}, \bibinfo{author}{Dong, S.S.}, \bibinfo{author}{Zhang, L.},
  \bibinfo{author}{Deng, F.Y.}, \bibinfo{author}{Zhao, Q.},
  \bibinfo{author}{Deng, H.W.}, \bibinfo{year}{2020}.
\newblock \bibinfo{title}{A road map for understanding molecular and genetic
  determinants of osteoporosis}.
\newblock \bibinfo{journal}{Nat. Rev. Endocrinol.} \bibinfo{volume}{16},
  \bibinfo{pages}{91--103}.
%Type = Article
\bibitem[{Yang et~al.(2017)Yang, Chen, Liu, Zhong, Qin, Lu, Feng and
  Chen}]{yang_cascade_2017}
\bibinfo{author}{Yang, W.}, \bibinfo{author}{Chen, Y.}, \bibinfo{author}{Liu,
  Y.}, \bibinfo{author}{Zhong, L.}, \bibinfo{author}{Qin, G.},
  \bibinfo{author}{Lu, Z.}, \bibinfo{author}{Feng, Q.}, \bibinfo{author}{Chen,
  W.}, \bibinfo{year}{2017}.
\newblock \bibinfo{title}{Cascade of multi-scale convolutional neural networks
  for bone suppression of chest radiographs in gradient domain}.
\newblock \bibinfo{journal}{Med. Image Anal.} \bibinfo{volume}{35},
  \bibinfo{pages}{421--433}.
%Type = Article
\bibitem[{Yu and Xia(2019)}]{yu_epidemiology_2019}
\bibinfo{author}{Yu, F.}, \bibinfo{author}{Xia, W.}, \bibinfo{year}{2019}.
\newblock \bibinfo{title}{The epidemiology of osteoporosis, associated
  fragility fractures, and management gap in {China}}.
\newblock \bibinfo{journal}{Arch. Osteoporos.} \bibinfo{volume}{14},
  \bibinfo{pages}{32}.
%Type = Inproceedings
\bibitem[{YUAN et~al.(2021)YUAN, Fu, Huang, Lin, Zhang, Chen and
  Wang}]{yuan_hrformer_2021}
\bibinfo{author}{YUAN, Y.}, \bibinfo{author}{Fu, R.}, \bibinfo{author}{Huang,
  L.}, \bibinfo{author}{Lin, W.}, \bibinfo{author}{Zhang, C.},
  \bibinfo{author}{Chen, X.}, \bibinfo{author}{Wang, J.}, \bibinfo{year}{2021}.
\newblock \bibinfo{title}{{HRFormer}: High-resolution vision {Transformer} for
  dense predict}, in: \bibinfo{booktitle}{NeurIPS}, pp.
  \bibinfo{pages}{7281--7293}.
%Type = Inproceedings
\bibitem[{Zhang et~al.(2020)Zhang, Zhang, Odena and
  Lee}]{zhang_consistency_2020}
\bibinfo{author}{Zhang, H.}, \bibinfo{author}{Zhang, Z.},
  \bibinfo{author}{Odena, A.}, \bibinfo{author}{Lee, H.}, \bibinfo{year}{2020}.
\newblock \bibinfo{title}{Consistency regularization for generative adversarial
  networks}, in: \bibinfo{booktitle}{ICLR}.
%Type = Inproceedings
\bibitem[{Zhao et~al.(2020)Zhao, Liu, Lin, Zhu and
  Han}]{zhao_differentiable_2020}
\bibinfo{author}{Zhao, S.}, \bibinfo{author}{Liu, Z.}, \bibinfo{author}{Lin,
  J.}, \bibinfo{author}{Zhu, J.Y.}, \bibinfo{author}{Han, S.},
  \bibinfo{year}{2020}.
\newblock \bibinfo{title}{Differentiable augmentation for data-efficient {GAN}
  training}, in: \bibinfo{booktitle}{NeurIPS}, pp. \bibinfo{pages}{7559--7570}.
%Type = Article
\bibitem[{Zhou et~al.(2021)Zhou, Canu, Vera and Ruan}]{zhou_latent_2021}
\bibinfo{author}{Zhou, T.}, \bibinfo{author}{Canu, S.}, \bibinfo{author}{Vera,
  P.}, \bibinfo{author}{Ruan, S.}, \bibinfo{year}{2021}.
\newblock \bibinfo{title}{Latent {Correlation} {Representation} {Learning} for
  {Brain} {Tumor} {Segmentation} {With} {Missing} {MRI} {Modalities}}.
\newblock \bibinfo{journal}{IEEE Transactions on Image Processing}
  \bibinfo{volume}{30}, \bibinfo{pages}{4263--4274}.
\newblock \DOIprefix\doi{10.1109/TIP.2021.3070752}.
%Type = Inproceedings
\bibitem[{Zhu et~al.(2017)Zhu, Park, Isola and Efros}]{zhu_unpaired_2017}
\bibinfo{author}{Zhu, J.Y.}, \bibinfo{author}{Park, T.},
  \bibinfo{author}{Isola, P.}, \bibinfo{author}{Efros, A.A.},
  \bibinfo{year}{2017}.
\newblock \bibinfo{title}{Unpaired {Image}-to-{Image} {Translation} {Using}
  {Cycle}-{Consistent} {Adversarial} {Networks}}, in: \bibinfo{booktitle}{2017
  {IEEE} {International} {Conference} on {Computer} {Vision} ({ICCV})}, pp.
  \bibinfo{pages}{2242--2251}.
\newblock \DOIprefix\doi{10.1109/ICCV.2017.244}. \bibinfo{note}{iSSN:
  2380-7504}.
%Type = Article
\bibitem[{Ziemlewicz et~al.(2015)Ziemlewicz, Binkley and
  Pickhardt}]{ziemlewicz_opportunistic_2015}
\bibinfo{author}{Ziemlewicz, T.J.}, \bibinfo{author}{Binkley, N.},
  \bibinfo{author}{Pickhardt, P.J.}, \bibinfo{year}{2015}.
\newblock \bibinfo{title}{Opportunistic {Osteoporosis} {Screening}: {Addition}
  of {Quantitative} {CT} {Bone} {Mineral} {Density} {Evaluation} to {CT}
  {Colonography}}.
\newblock \bibinfo{journal}{J. Am. Coll. Radiol.} \bibinfo{volume}{12},
  \bibinfo{pages}{1036--1041}.
%Type = Article
\bibitem[{Zou et~al.(2020)Zou, Liu, Cao, Liu, He, Peng and
  Shuai}]{zou_advances_2020}
\bibinfo{author}{Zou, Z.}, \bibinfo{author}{Liu, W.}, \bibinfo{author}{Cao,
  L.}, \bibinfo{author}{Liu, Y.}, \bibinfo{author}{He, T.},
  \bibinfo{author}{Peng, S.}, \bibinfo{author}{Shuai, C.},
  \bibinfo{year}{2020}.
\newblock \bibinfo{title}{Advances in the occurrence and biotherapy of
  osteoporosis}.
\newblock \bibinfo{journal}{Biochem. Soc. Trans.} \bibinfo{volume}{48},
  \bibinfo{pages}{1623--1636}.

\end{thebibliography}

\end{document}

\verso{Yi Gu \textit{et~al.}}

\begin{frontmatter}

\title{SUPPLEMENTAL MATERIALS \newline Bone mineral density estimation from a plain x-ray image by learning decomposition into projections of bone-segmented computed tomography}

% \author[1,4]{Yi \snm{Gu}\corref{cor1}}
% \ead{gu.yi.gu4@is.naist.jp}
% \cortext[cor1]{Corresponding authors at ICB lab., Division of Information Science, Graduate School of Science and Technology, Nara Institute of Science and Technology, Japan.}
% \author[1]{Yoshito \snm{Otake}\corref{cor1}}
% \ead{otake@is.naist.jp}

% \author[2]{Keisuke \snm{Uemura}}

% \author[1]{Mazen \snm{Soufi}}

% \author[3]{Masaki \snm{Takao}}

% \author[4]{Hugues \snm{Talbot}}

% \author[2]{Seiji \snm{Okada}}

% \author[5]{Nobuhiko \snm{Sugano}}

% \author[1]{Yoshinobu \snm{Sato}\corref{cor1}}
% \ead{yoshi@is.naist.jp}

\end{frontmatter}

%\linenumbers


\newcommand{\rmtxt}[1]{\textcolor{red}{\sout{#1}}}
\newcommand{\adtxt}[1]{{#1}}
% \pagebreak
\begin{figure*}[h]
\centering
\includegraphics[width=\textwidth]{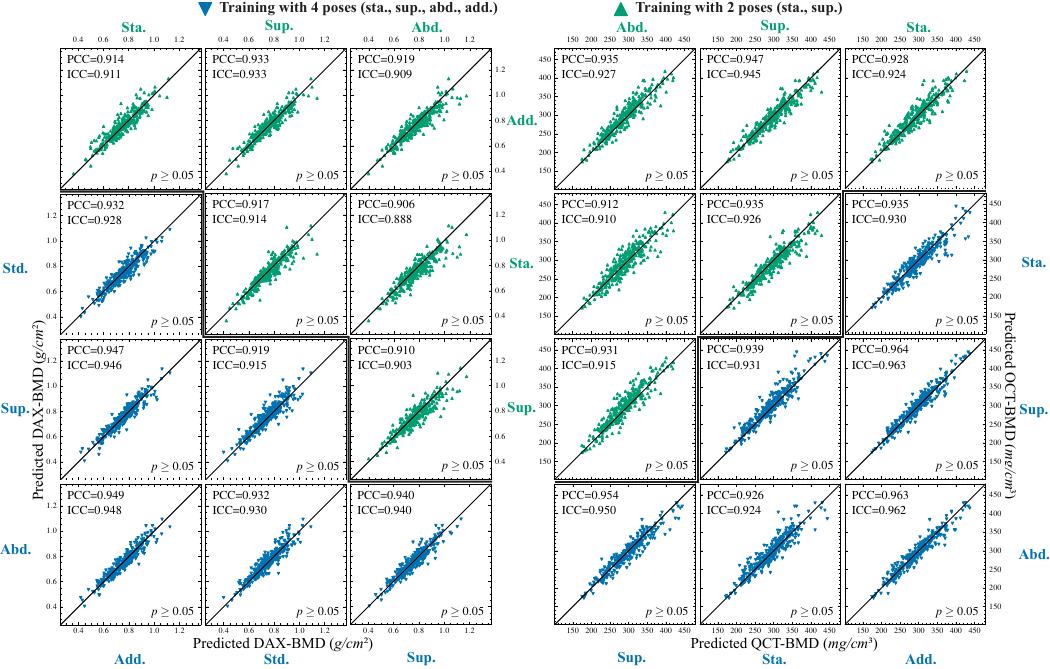}
\caption{Correlations of predicted BMD between poses, including std. (standing), sup. (supine), abd. (abduction), and add. (adduction), obtained by the proposed method on the experiments using (blue scatters) 4 poses and (green scatters) 2 poses in training.
Statistical significance was denoted by $p$-value, which was computed using Tukey honest significant differences test, where the difference between two distributions with $p>0.05$ was considered non-significant.}
\label{fig:315_pose_to_pose_bmd}
\end{figure*}

\begin{figure*}[!t]
\centering
\includegraphics[width=\textwidth]{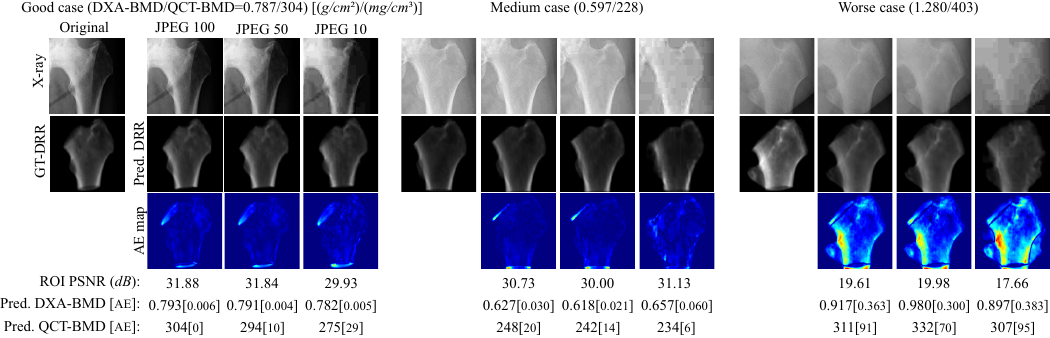}
\caption{Sampled good, medium, and worse cases with respect to the BMD estimation accuracy in compression experiment, showing the ROI of original x-ray images, compressed x-ray images, ground-truth (GT) DRR, predicted (Pred.) DRR in stage 2, the absolute error (AE) map between pred. and GT DRR.}
\label{fig:compress_representative}
\end{figure*}

\begin{figure*}[!t]
\centering
\includegraphics[width=\textwidth]{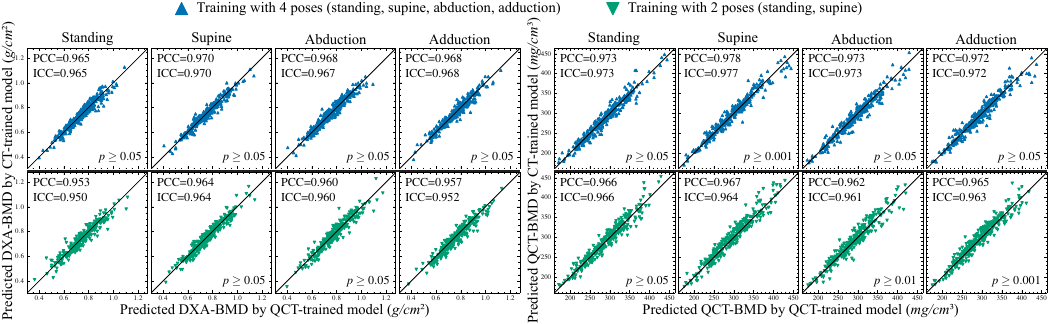}
\caption{Correlations of predicted BMD by the proposed method between training with QCT and uncalibrated CT grouped by the poses used in training.
Statistical significance was indicated by $p$-values, where $p$ larger than 0.05 were considered non-significant.}
\label{fig:315_quant_vs_uncalib}
\end{figure*}

\begin{figure*}[!t]
\centering
\includegraphics[width=\textwidth]{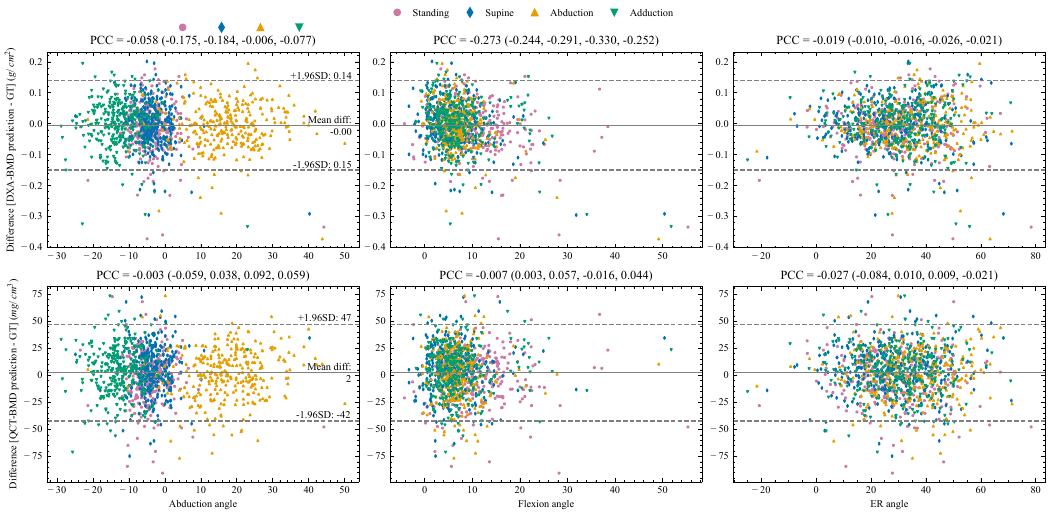}
\caption{\adtxt{Scatters of hip angles against BMD prediction error for error source investigation. The Pearson correlation coefficients (PCCs) for all poses, standing pose, supine pose, abduction pose, and adduction pose are shown at the top of each scatter. Though a weak trend (PCC=-0.273) was observed between flexion angle and DXA-BMD prediction error, no strong correlation appeared overall, likely because of the noise on the femoral head deformation caused by diseases in our dataset.}}
\label{fig:angle_vs_error}
\end{figure*}

\begin{figure*}[!t]
\centering
\includegraphics[width=\textwidth]{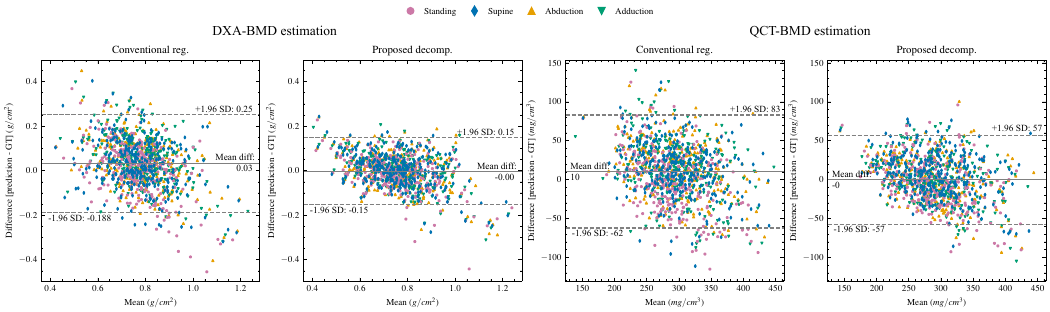}
\caption{\adtxt{Bland-altman plots for the conventional reg. and proposed decomp. methods on the (left) DXA-BMD and (right) QCT-BMD estimation tasks.}}
\label{fig:methods_bland_altman}
\end{figure*}

\begin{figure*}[!t]
\centering
\includegraphics[width=\textwidth]{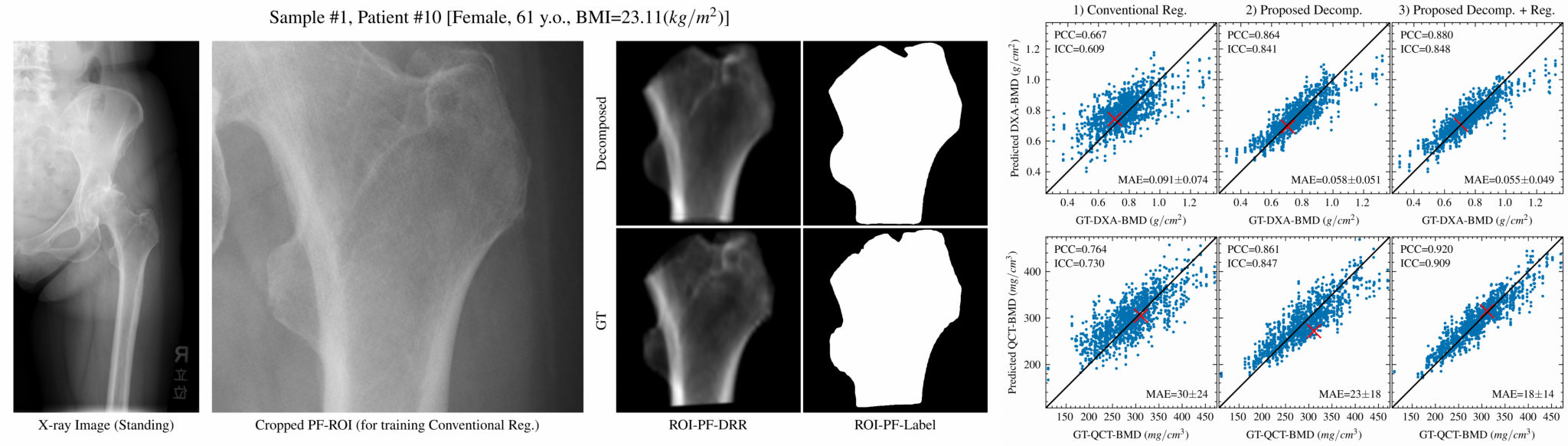}
\caption{\adtxt{Screenshot of visualization video of 1206 samples (305 patients), visualizing the hemi-hip X-ray image, the ROI X-ray image (used to train conventional methods), the ROI-PF-DRR, the ROI-PF label., and BMD estimation results.}}
\label{fig:315_screenshot}
\end{figure*}

\begin{figure*}[!t]
\centering
\includegraphics[width=\textwidth]{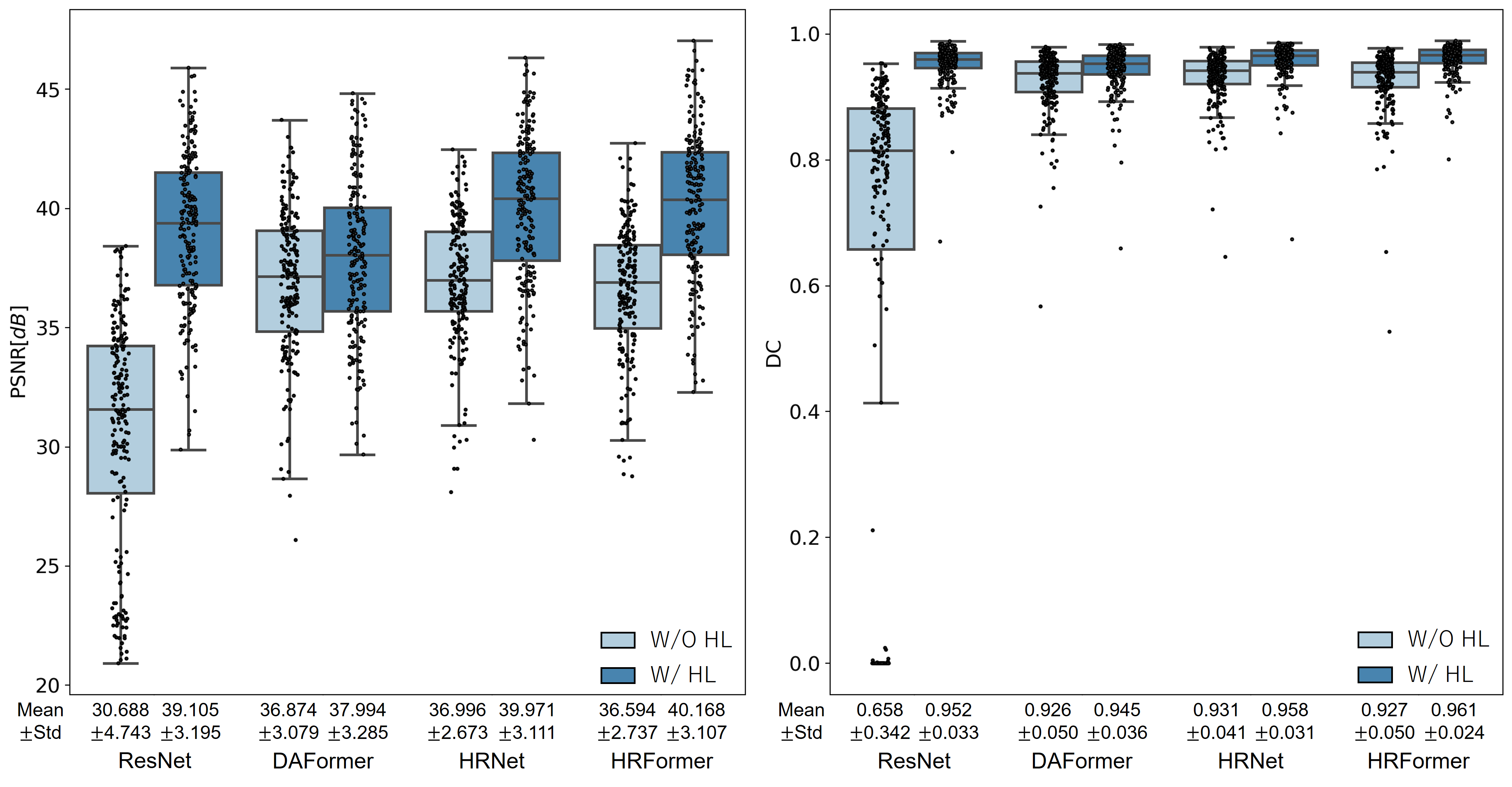}
\caption{Quantitative evaluation results of image decomposition for different backbones w/o and w/ hierarchical learning (HL). 
The performance improvements by HL were observed for all the backbones, where the ResNet backbone benefited the most from HL. 
The lowest accuracy among the methods with HL (DAFormer w/ HL) was higher than the best accuracy among the methods without HL (DAFormer w/o HL). 
The comparison between HRNet and HRFormer suggests that HRNet slightly outperformed the HRFormer under no HL, while applying HL boosted both of them and made HRFormer a better choice.}
\label{fig:315_quant_vs_uncalib}
\end{figure*}

\begin{figure*}[!t]
\centering
\includegraphics[width=\textwidth]{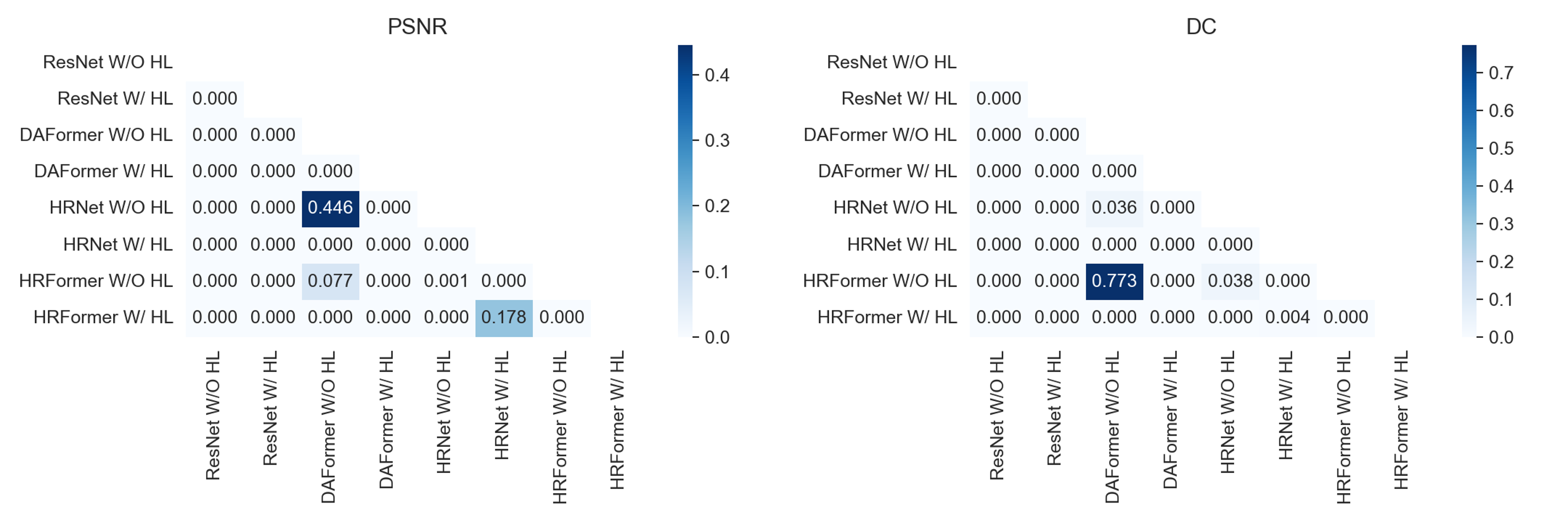}
\caption{Results of the statistical test on PSNR and DC between methods. 
The $p$ values between the methods without HL and with HL were smaller than 0.001 for all backbones, indicating significant differences. 
The test results showed the DAformer, HRNet, and HRFormer had no significant difference when HL was not used, whereas the differences were more pronounced when HL was used.}
\label{fig:315_quant_vs_uncalib}
\end{figure*}

\begin{figure*}[!t]
\centering
\includegraphics[width=\textwidth]{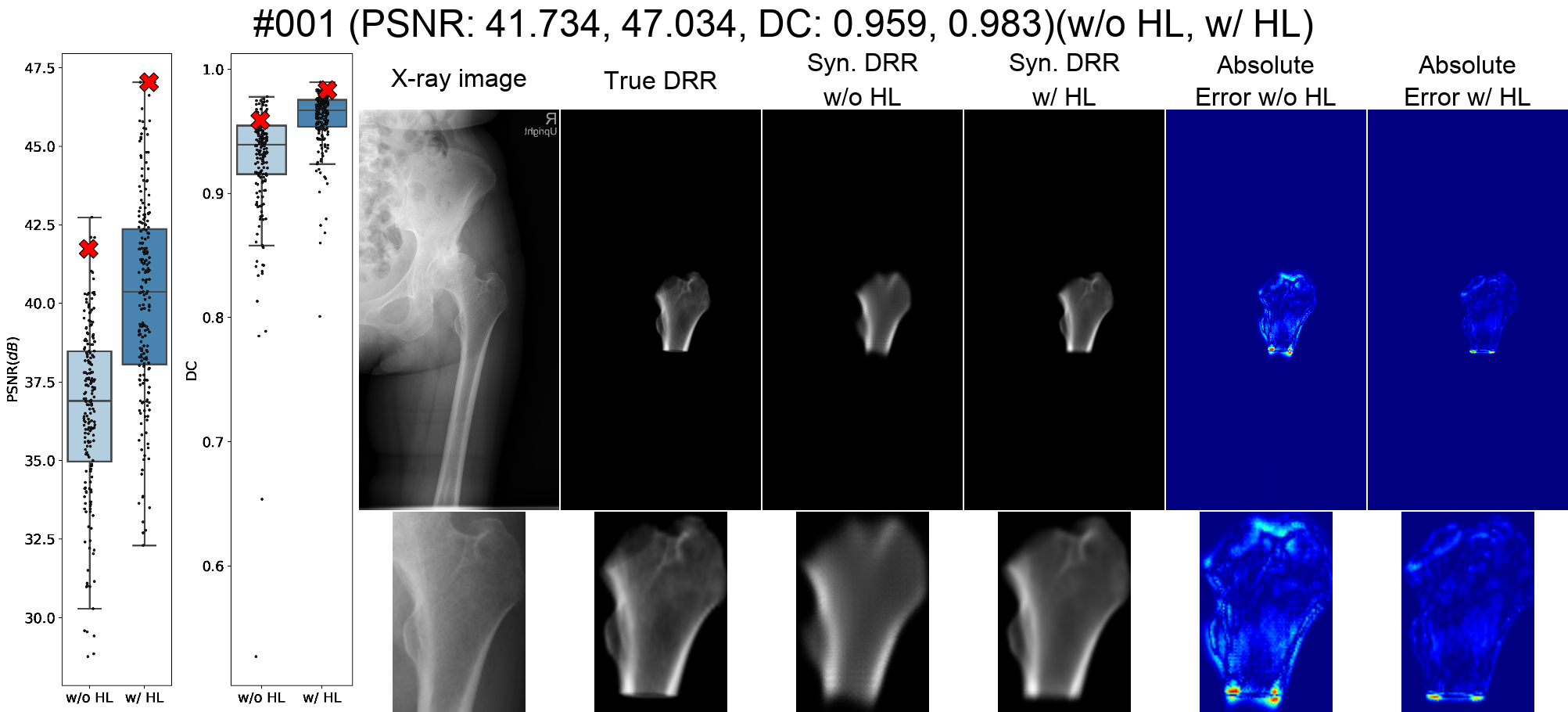}
\caption{Screenshot of visualization video of 200-cases, visualizing decomposition results by HRFormer w/o and w/ HL in the descending order of PSNR by HRFormer w/ HL, where the more consistent shape and more detailed inner structure by applying HL were observed visually. 
The absolute error maps suggest the significant error is more likely produced by the upper and bottom cutting edges, which are clinically defined while applying HL made the model better in handling the cutting edges.}
\label{fig:315_quant_vs_uncalib}
\end{figure*}